\documentclass[superscriptaddress,onecolumn]{revtex4}
\usepackage{graphicx,epsfig}
\usepackage{amsmath}
\usepackage {amssymb}
\usepackage{multirow}
\usepackage{float}
\usepackage{color}
\providecommand{\abs}[1]{\lvert#1\rvert}

\newcommand{\be}{\begin{eqnarray}}
\newcommand{\ee}{\end{eqnarray}}
\newcommand{\bea}{\begin{eqnarray}}

\newcommand{\eea}{\end{eqnarray}}

\makeindex

\begin{document}

\title{ Massive Scalar Field Perturbations of Black Holes Surrounded by Dark Matter}

\author{Ram\'{o}n B\'{e}car}
\email{rbecar@uct.cl}
\affiliation{Departamento de Ciencias Matem\'{a}ticas y F\'{\i}sicas, Universidad Catolica de Temuco}

\author{P. A. Gonz\'{a}lez}
\email{pablo.gonzalez@udp.cl} \affiliation{Facultad de
Ingenier\'{i}a y Ciencias, Universidad Diego Portales, Avenida Ej\'{e}rcito
Libertador 441, Casilla 298-V, Santiago, Chile.}

\author{Eleftherios Papantonopoulos}
\email{lpapa@central.ntua.gr}
\affiliation{Department of Physics, National Technical University of Athens, Zografou Campus GR 157 73, Athens, Greece.}

\author{Yerko V\'asquez}
\email{yvasquez@userena.cl}
\affiliation{Departamento de F\'isica, Facultad de Ciencias, Universidad de La Serena,\\
Avenida Cisternas 1200, La Serena, Chile.}

\date{\today}

\begin{abstract}
We consider scalar field perturbations in the background of black holes immersed in perfect fluid dark matter (PFDM). We find, by using the sixth order Wentzel-Kramers-Brillouin (WKB) approximation, that
the longest-lived modes are the ones with higher angular number,
for a scalar field mass smaller than
a critical value, known as anomalous decay rate of the quasinormal modes, while that beyond this critical value the behaviour is inverted. Moreover, we show that it is possible to recover the real part of the quasinormal frequencies (QNFs), the imaginary part of the QNFs, and the critical scalar field mass, of the Schwarzschild background for different values of the PFDM intensity parameter $k$, respectively. For values of $k$ smaller than these values, the mentioned quantities are greater than the Schwarzschild background. However, beyond of these values of $k$, these quantities are smaller than the Schwarzschild background.

\end{abstract}

\maketitle

\tableofcontents

%\begin{document}
%\maketitle
%\flushbottom
\clearpage

%\begin{document}
\maketitle
\flushbottom

\section{Introduction}

The recent observations of gravitational waves  (GWs) produced as a result of relativistic collision of
two compact objects, opened up the possibility of studying the nature and the properties of compact objects. In the near future, following the recent LIGO detections \cite{Abbott:2016blz}-\cite{TheLIGOScientific:2017qsa}, GW astronomy will provide us a better understanding of the gravitational interaction and astrophysics of these objects. The recent observations do not yet probe the detailed structure of spacetime of compact objects
beyond the photon sphere; however, one expects that the properties of these objects will be revealed in the years to come with future GW observations. In particular, the expectation is to precisely detect the ringdown phase, which is governed by a series of damped oscillatory
modes at early times, named quasinormal modes (QNMs) \cite{Vishveshwara:1970zz}-\cite{Konoplya:2011qq}. It is expected that the future GW observations will give  some information on the nature and physics of the near-horizon region of black holes (BHs). The existence of any
structure at near-horizon scales would generate a series of “echoes” of the primary gravitational
wave signal, produced during the ringdown phase \cite{Cardoso:2016rao,Cardoso:2016oxy}.

In General Relativity one of the most important compact object is the  BH. One of the most distinguishable feature of BHs is the  event horizon. The event horizon which is a causal boundary, does not allow us to go inside the BH. Black holes in our Universe may be affected by astronomical
environments, such as the dark matter (DM) near the BHs \cite{Ferrer}-\cite{Xu}. It is believed that $90\%$   of galaxies
 are composed of DM \cite{Jusufi}. The abnormally high velocities of stars at the outskirts of galaxies imply that visible disks of galaxies are immersed in a much larger roughly spherical halo of DM \cite{Kafle:2014xfa,Battaglia:2005rj}. The DM does not interact with the electromagnetic field and therefore the propagation of light is possible inside the dark matter halo. Some studies to learn whether the black hole shadow could be affected by the tidal forces induced by the invisible matter have been performed in \cite{Hou:2018avu,Haroon:2018ryd,Hou:2018bar}.
However, the results look highly model-dependent, due to a particular equation of state for the dark matter was considered.
 Then the most reliable way to detect dark matter outside the BH is to study the gravitational wave signal, produced during the ringdown phase
 generating a series of echoes. Various models of black holes surrounded by dark matter have been studied \cite{Jusufi:2022jxu}-\cite{Xavier:2023exm}.

On the other hand, recent astrophysical observations  show an accelerating expansion of Universe \cite{Perl}, implying the presence of a state with
negative pressure. The negative pressure could be originated from the presence of  barotropic perfect fluid which corresponds to the dark energy or to the presence of a cosmological constant. The state equation is given by the relation between the pressure
$p$ and the energy density $\rho$, $w=p/\rho$ and the recent observations suggest that  the equation of state lies in a narrow strip around $w =-1$ \cite{WMAP:2010qai,Alam:2003fg}, where  $w =-1$  corresponds to a cosmological constant $\Lambda$ and $w <-1$ is allowed \cite{Caldwell:1999ew,Caldwell:2003vq}. Considering that the Universe is filled with a barotropic perfect fluid corresponding to dark energy, an equation of state $w <-1$ can be realized with the presence of a phantom field. The phantom dark energy has the   property that the dominant energy condition is violated so  that the energy density and curvature may grow to infinity in a finite time, which is referred to a Big Rip singularity \cite{Caldwell:2003vq,Nesseris:2004uj}. The introduction of dark matter may explain the discrepancy between the predicted rotation curves of galaxies when only including luminous matter and the actual (observed) rotation  curves which differ significantly \cite{deAlmeida:2018kwq,Harada:2022edl,Shabani:2022buw}.

A spherical halo of dark matter around the  Schwarzschild BH has been considered for testing the gravitational response
%(such as quasinormal modes \cite{Konoplya:2011qq} or echoes \cite{Cardoso:2017cqb})
of black holes in the astrophysical environment \cite{Barausse:2014tra,Leung:1997was,Konoplya:2018yrp}, and also to study the shadow \cite{Konoplya:2019sns}. The dark matter possesses some mass which can be modelled as an additional effective mass in the mass function of a BH.
Then one expects that the ringdown profile to be   modified due to new physics near the surface/event horizon and echoes to be generated due
to matter at some distance from the black hole. This allows
us to understand how the echoes of the surface of the BH are affected by the astrophysical
environment at a distance. We expect that the straightforward calculations for the time-domain profiles of such
a system to support the expectations that if the echoes are observed, they should most probably be
ascribed to some new physics near the event horizon rather than some “environmental” effect.  An exact and fully relativistic solution of the Einstein field equations was discussed  in \cite{Cardoso:2021wlq}  which describes a black hole in the center of an anisotropic  dark matter halo without the need of adding any Newtonian potential for the  dark matter, and  metric perturbations in these spacetimes were also studied \cite{Cardoso:2022whc}.           

Static spherically-symmetric exact solutions of Einstein equations with the quintessential matter having negative equation of state surrounding a black hole charged or not was presented in \cite{Kiselev:2002dx,Kiselev:2003ah}. A condition of additivity and linearity in the energy-momentum tensor was introduced, allowing to get the  known solutions for the electromagnetic static field,  implying a relativistic relation between the energy density and
pressure. A real phantom field minimally coupled to gravity was introduced in \cite{Li:2012zx}, with the aim to investigate the possibility that the galactic dark matter exists in an scenario with a phantom field responsible for the dark energy, where there is a static and spherical approximate solution for this kind of the galaxy system with a supermassive black hole at its center. The phantom field defined in an effective theory is valid only at low energies, in order to avoid the well-known quantum instability of the vacuum at high frequencies.

The aim of this work is to study the propagation of massive scalar fields in a BH spacetime surrounded by perfect fluid dark matter (PFDM) for highest values of $\ell$ in order to see  if there is an anomalous decay rate of QNMs  for the photon sphere modes \cite{Cardoso:2008bp}. 
The QNMs give an infinite discrete spectrum which consists of complex frequencies, $\omega = \omega_R + i\omega_I$, where the real part $\omega_R$ determines the oscillation timescale of the modes, while the complex part $\omega_I$ determines their exponential decaying timescale. If the background consists  of Schwarzschild and Kerr
BHs it was found that for gravitational perturbations the longest-lived modes are always the ones with lower angular number $\ell$.
This is understood from the fact that the more energetic modes with high angular number $\ell$ would have faster
decaying rates. The anomalous behaviour occurs when the longest-lived modes are the ones with higher angular number and this can occurs  with a massive probe scalar field. There is a critical mass of the scalar field  where the behaviour of the decay rate of the QNMs is inverted and can be obtained from the condition $Im(\omega)_{\ell}=Im(\omega)_{\ell+1}$ in the {\it eikonal} limit, that is when $\ell \rightarrow \infty$. The anomalous behaviour in the quasinormal frequencies (QNFs) is possible in asymptotically flat, in asymptotically dS and
 in asymptotically AdS spacetimes; however, we observed that the critical mass exist for asymptotically flat and for asymptotically dS spacetimes and it is not present in asymptotically AdS spacetimes for large and intermediate BHs. Moreover, such behaviour have been studied for scalar fields \cite{Konoplya:2004wg, Konoplya:2006br,Dolan:2007mj, Tattersall:2018nve,Lagos:2020oek, Aragon:2020tvq,Aragon:2020xtm,Fontana:2020syy,Becar:2023jtd,Gonzalez:2022ote} as well as charged scalar fields \cite{Gonzalez:2022upu,Becar:2022wcj} and fermionic fields \cite{Aragon:2020teq}.  The  anomalous decay in accelerating black holes was studied in \cite{Destounis:2020pjk}.

The QNMs for a massless  scalar field and their connection to the shadow radius in the background of a modified Schwarzschild BH was studied in \cite{Jusufi:2019ltj}. The modification of the background BH was due to the presence of the PFDM  surrounding  the BH  encoded by the parameter $k$.
It was shown  that the QNM spectra  deviates from those of Schwarzschild BH due to the presence of the PFDM. Moreover, for any $k > 0$, the real part and the absolute value of the imaginary part of QNFs
%QNM frequencies 
increases and this means that the field perturbations in the presence of PFDM decays more rapidly compared to Schwarzschild vacuum BH. Also, it was shown that there exists a reflecting point $k_0$ corresponding to maximal values for the real part of QNM frequencies. Namely, as the PFDM parameter $k$ increases in the interval $k < k_0$, the QNFs
increases and reach their maximum values at $k = k_0$. Also, it was shown that  $k_0$ is also a reflecting point for the shadow radius. In the physical context, QNFs
are associated with the value of the effective potential and its derivatives evaluated at its maximum point. For asymptotically flat black holes, this radius corresponds to the radius of the photon sphere or the radius of unstable null geodesics. Furthermore, as shown in \cite{Cardoso:2008bp}, utilizing the WKB approximation and establishing the connection between unstable null geodesics and QNFs
%quasinormal frequencies 
in the eikonal limit, the following  relation was found for the real part of the QNFs 
    $\omega_{real}=l\sqrt{\frac{f(r_{c})}{r_{c}^{2}}}$,
where $r_{c}$ is the radius of unstable null geodesics or the photon sphere.  
Additionally, it was demonstrated in \cite{Stefanov:2010xz,Jusufi:2019ltj,Cuadros-Melgar:2020kqn,Anacleto:2021qoe} that the black hole shadow radius can be related to the real part of QNFs
in the eikonal limit:
    $\omega_{real}=\frac{l}{R_{s}}$,
where $R_{s}=\frac{r_{c}}{\sqrt{f(r_{c})}}$.
As indicated in the article \cite{Jusufi:2019ltj}, the shadow radius attains its lowest value for $k=k_{0}=0.81$, which, according to the expression for $\omega_{real}$, 
corresponds to the largest quasinormal oscillation (real part). This aligns with the results obtained through the WKB method. It is worth noting that if the event horizon decreases, it will lead to a reduction in the shadow radius and, consequently, higher values for the oscillation frequencies.

However, the analysis in \cite{Jusufi:2019ltj}  was carried out for low values of $\ell$,  and it is known that the WKB method provides better accuracy for larger $\ell$ (and $\ell > n$ where $n$ is the overtone). In this study, we consider larger values of $\ell$, and we show that there is a value of the parameter $k$, for which the real part of the QNFs is maximum, which is associated with a reflecting point in the context of BH shadow.
%there exists a reflecting point.
Additionally, our results show that there is a value of the parameter $k$, for which the absolute value of the imaginary part of the QNFs is maximum. As we will see, there is a value of $k \neq 0$, that reproduces the real and the imaginary part of the QNFs of the Schwarzschild background separately, but there is not a value of $k$ that can reproduce the same QNF that the Schwarzschild background. Also, we show the existence of an anomalous behaviour of the decay rate for the background considered. Moreover, there is a value of $k \neq 0$, that reproduces the critical scalar field mass of the Schwarzschild background.

The work is organized as follows. In Sec. \ref{setup}, we give the setup of the theory.
%we considered.
Then, in Sec. \ref{perturbations} we study the scalar perturbations, and in Sec. \ref{WKBJ} we study the photon sphere modes. Finally, conclusions and final remarks are presented in Sec. \ref{conclusion}.

\section{The setup of the theory}
\label{setup}

One of the first works that discussed static spherically-symmetric exact solutions of Einstein
equations with the dark matter surrounding a black hole was discussed in \cite{Kiselev:2002dx}.
A condition of additivity and linearity in the energy-momentum tensor was introduced, which
allows one to get correct limits to the known solutions implying the relativistic relation between the energy density and
pressure. The solution which was found is
\begin{equation}
{\rm d}s^2 = \left[1-\frac{r_g}{r}-\sum_n \left( \frac{r_n}{r} \right)^{3w_n+1}
\right]\,{\rm d}t^2-\frac{{\rm d}r^2}{\displaystyle\left[1-\frac{r_g}{r}-\sum_n
\left(\frac{r_n}{r} \right)^{3w_n+1} \right]} - r^2({\rm d}\theta^2 -
\sin^2\theta\, {\rm d}\phi^2)~,
\label{general solution}
\end{equation}
where $r_g = 2 M$, $M$ is the black hole mass, $r_n$ are the dimensional
normalization constants, and $w_n$ are the dark matter state parameters
\begin{equation}
p_n=w_n \rho_n~.
\end{equation}

This work was further extended in \cite{Kiselev:2003ah} in which a scalar field was introduced. Then the state describing the dark matter
 with the negative pressure was
considered as a perfect fluid approximation of a scalar field with
an appropriate potential. So, introducing a scalar field $\varphi$ with the Lagrangian equal to
\begin{equation}\label{lag0}
    {\cal L} = \frac{1}{2}\, g^{\mu\nu}\,\partial_\mu \varphi\,\partial_\nu \varphi
    -V(\varphi)~,
\end{equation}
with identification
$$
\rho_{DM} = - g^{tt}[\partial_r \varphi(r)]^2,
$$
the dark matter modified metric (\ref{general solution}) was reproduced.

The possibility that a phantom field is responsible for the dark matter was further investigated in \cite{Li:2012zx}. A spherically approximate solution of a dark matter galaxy was obtained with a  supermassive black hole at its center. The solution of the metric functions is satisfied with $g_{tt} = - g_{rr}^{-1}$ and   the observation of the rotational stars moving in circular orbits  in a spiral galaxy constrained  the background of the phantom field  to be spatially inhomogeneous having an exponential potential.

The action of real phantom field minimally coupled to gravity
was given by
\begin{equation}
S=\int d^4x\sqrt{-g}\left[\frac{R}{2\kappa ^2}+\frac{1}{2}g^{uv}\partial_{u}\Phi\partial_{v}\Phi-V(\Phi)+\mathcal{L}_{m}\right]~,
\label{action}
\end{equation}
where $R$ is the Ricci scalar, $ {\kappa ^2} = 8 {\pi G}$ being $G$ the Newton constant, and $V(\phi)$ is the phantom field potential. The term $\mathcal{L}_{m}$ account for the massive dark matter in the galaxy.

 The black hole solution is described by the metric
\begin{equation}
    ds^2=-f(r)dt^2+\frac{dr^2}{f(r)}+r^2(d\theta^2+sin^2\theta d\phi^2)\,,
    \label{metric}
\end{equation}
where
\begin{equation}
f(r)=1-\frac{2M}{r}+\frac{k}{r}\ln(\frac{r}{\abs{k}})\,,\label{metfun}
\end{equation}
$M$ is the black hole mass and $k$ is a parameter describing the intensity of the PFDM. The above metric function is similar to the metric function with the parameter $k$ specifying the presence of the phantom scalar field acting as its scalar charge. It was found that because the infalling phantom particles have a total negative energy,  the accretion of the phantom energy is related to the decrease of the black hole mass.

Using (\ref{metric}) as the background metric   with the metric function (\ref{metfun}), massless scalar field and electromagnetic field perturbations were carried out in \cite{Jusufi:2019ltj}. It was found that the presence of the PFDM  parameter $k$ results in a deviation of  quasinormal mode spectra from those of Schwarzschild black hole and it was shown that the field perturbations in the presence of PFDM decay more rapidly compared to Schwarzschild vacuum black hole.

\section{Scalar field perturbations}
\label{perturbations}

\begin{figure}[h!]
\includegraphics[width=0.4\textwidth]{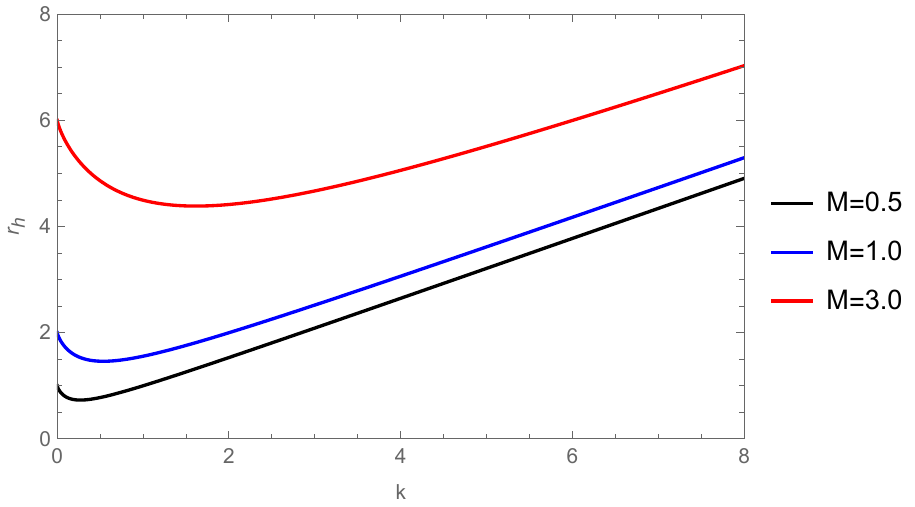}
\caption{The behaviour of the event horizon radius $r_h$ as a function of the PFDM intensity parameter $k$. Black line for $M=0.5$, blue line for $M=1.0$, and red line for $M=3.0$.}
\label{F9}
\end{figure}

We first show in Fig. \ref{F9}, the behaviour of the event horizon  radius $r_h$ of the BH as a function of the parameter $k$ (in the following we consider $k > 0$) using the metric function (\ref{metfun}). We can observe that when the BH mass $M$ increases, the event horizon  radius increases. Also, the event horizon is minimum for $k=k_h=\frac{2M}{1+e}$.
%there is a value of $k_h=\frac{2M}{1+e}$ with$ \textbf{ELIMINAR} e>0$, corresponding to a minimal value for the event horizon.
Namely, as the PFDM parameter $k$ increases in the interval $k < k_h$, $r_h$ decreases and reaches  its minimum value at $k = k_h$.
Then, for $k > k_h$, $r_h$  increases when the parameter $k$ increases.
It is worth noting that there is a value of $k$, for which the event horizon  radius is the same  as the  Schwarzschild BH, and beyond this value of $k$ the event horizon is greater than the Schwarzschild background.
The QNMs of scalar perturbations in the background of the metric (\ref{metric})
are given by the scalar field solution of the Klein-Gordon equation
\begin{equation}
\frac{1}{\sqrt{-g}}\partial _{\mu }\left( \sqrt{-g}g^{\mu \nu }\partial_{\nu } \varphi \right) =m^{2}\varphi \,,  \label{KGNM}
\end{equation}%
with suitable boundary conditions for a BH geometry. In the above expression $m$ is the mass
of the scalar field $\varphi $. Now, by means of the following ansatz
\begin{equation}
\varphi =e^{-i\omega t} R(r) Y(\Omega) \,,\label{wave}
\end{equation}%
the Klein-Gordon equation reduces to
\begin{equation}
\frac{1}{r^2}\frac{d}{dr}\left(r^2 f(r)\frac{dR}{dr}\right)+\left(\frac{\omega^2}{f(r)}-\frac{\kappa^{2}}{r^2}-m^{2}\right) R(r)=0\,, \label{radial}
\end{equation}%
where we defined $-\kappa^{2}=-\ell (\ell+1)$, with $\ell=0,1,2,...$, which represents the eigenvalues of the Laplacian on the two-sphere and $\ell$ is the multipole number.
Now, defining $R(r)=\frac{F(r)}{r}$
and by using the tortoise coordinate $r^*$ given by
$dr^*=\frac{dr}{f(r)}$,
 the Klein-Gordon equation can be written as a one-dimensional Schr\"{o}dinger-like equation
 \begin{equation}\label{ggg}
 \frac{d^{2}F(r^*)}{dr^{*2}}-V_{eff}(r)F(r^*)=-\omega^{2}F(r^*)\,,
 \end{equation}
 with an effective potential $V_{\text{eff}}(r)$, which  parametrically thought,  $V_{eff}(r^*)$, is given  by
  \begin{equation}\label{pot}
 V_{eff}(r)=\frac{f(r)}{r^2} \left(\kappa^{2} + m^2 r^2+f^\prime(r)r\right)\,.
 \end{equation}

\begin{figure}[h!]
\begin{center}
\includegraphics[width=0.3\textwidth]{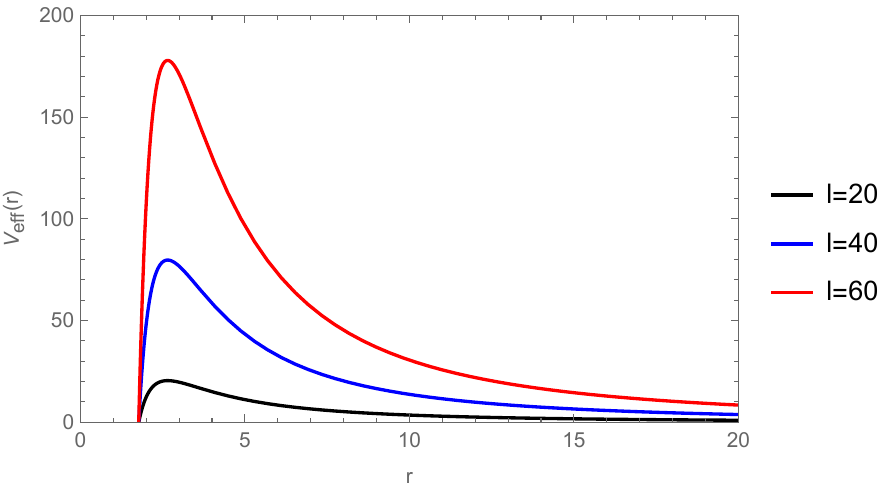}
\includegraphics[width=0.3\textwidth]{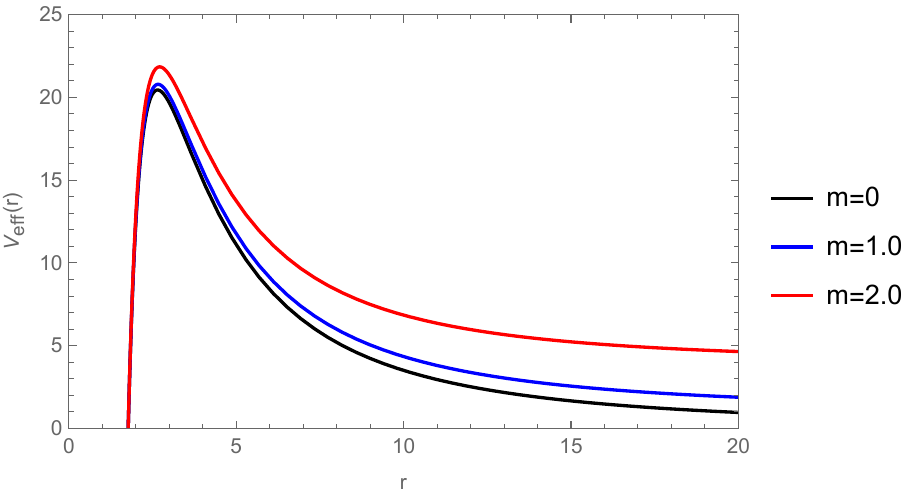}
\includegraphics[width=0.3\textwidth]{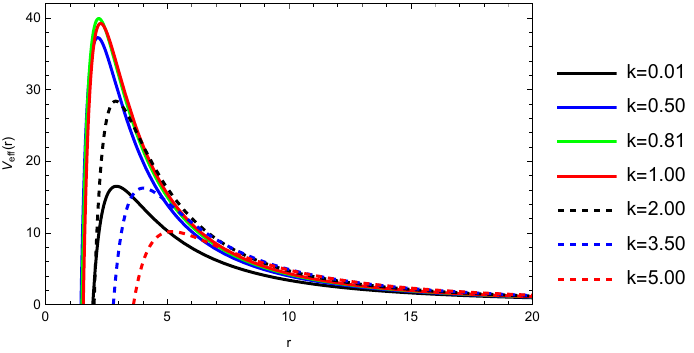}
\end{center}
\caption{The effective potential $V_{eff}$ as a function of $r$, with $M=1$. Left panel for massless scalar field $m=0$, $k=0.01$, and $\ell=20,40,60$. Central panel for $\ell=20$, $k=0.01$, and  $m=0,1.0,2.0$. Right panel for $\ell=20$, $m=0$, and $k=0.01, 0.50, 1.00, 2.00, 3.50, 5.00$.}
\label{Potential}
\end{figure}

 In Fig. \ref{Potential}, we show  its behaviour  for different values of the parameters. Note that a barrier of potential occurs  in each case, with  its maximum value increasing when $\ell$ or $m$ increases. However, for $0<k \leq 0.81$ the maximum in the potential increases when $k$ increases. But for $k>0.81$, the maximum  of the potential decreases when $k$ increases, being possible to recover the  height of the potential for the Schwarzschild case ($k\approx 3.50$), but with a  larger event horizon.

\newpage

\begin{figure}[h]
\begin{center}
\includegraphics[width=0.32\textwidth]{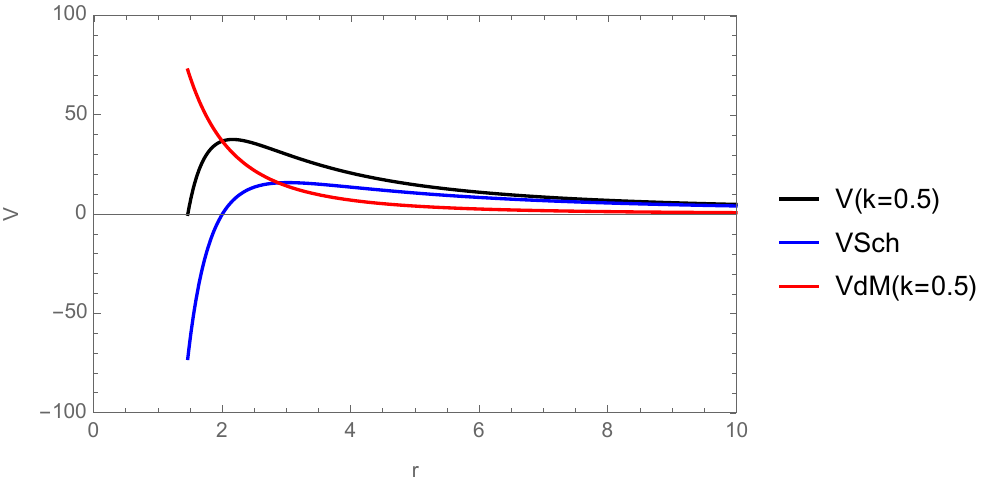}
\includegraphics[width=0.32\textwidth]{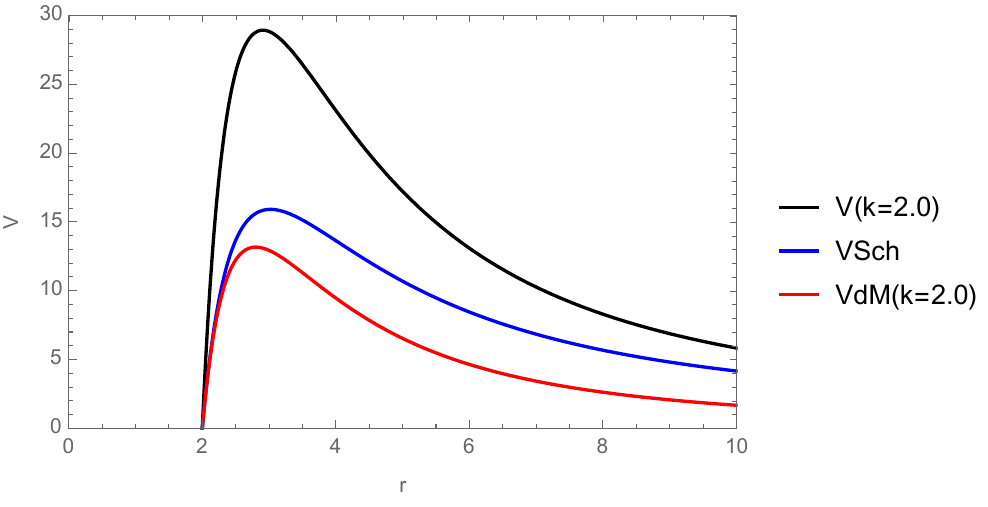}
\includegraphics[width=0.32\textwidth]{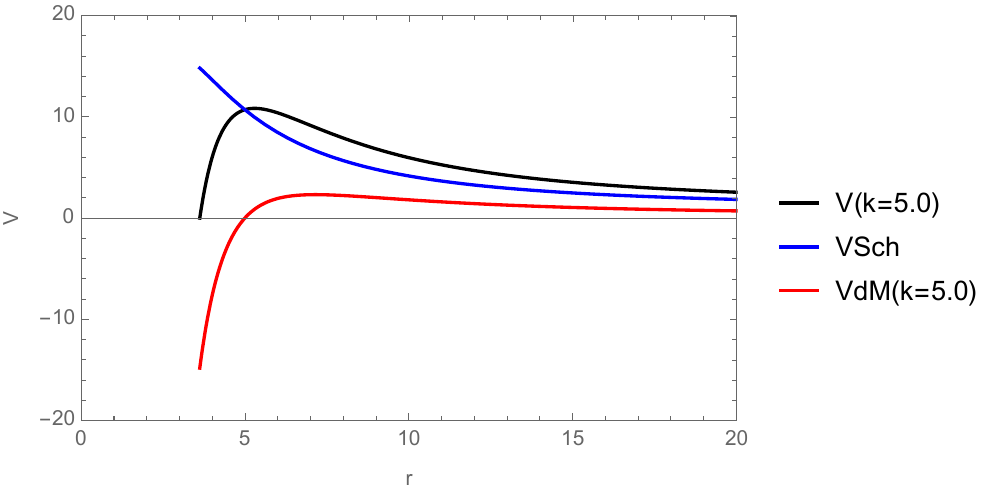}
\end{center}
\caption{The behaviour of the potentials $V_{Sch}$, $V_{dM}$ and $V=V_{Sch}+V_{dM}$  as a function of $r$ with $m=1$, $M=1$, and $\ell=20$. Black line for $V$, blue line for $V_{Sch}$, and red line for $V_{dM}$. Left panel for $k=0.5$, and $r_h=1.463$, central panel for $k=2.0$, and $r_h=2$, and right panel for $k=5.0$, and $r_h=3.618$.}
\label{PotentialS}
\end{figure}

The effective potential can be  splitted as $V_{eff}(r)= V_{Sch}(r) + V_{dM}(r)$, where
\begin{equation}
\label{VSch}
    V_{Sch}(r)=
    %-\frac{2 l^2 M}{r^3}+\frac{l^2}{r^2}-\frac{2 l M}{r^3}+\frac{l}{r^2}-\frac{2 m^2 M}{r}+m^2-\frac{4 M^2}{r^4}+\frac{2 M}{r^3}=
    -\frac{(2 M-r) \left(l (l+1) r+m^2 r^3+2 M\right)}{r^4}\,,
\end{equation}

\begin{equation}
\label{VdM}
     V_{dM}(r)=
     %-\frac{k^2 \log ^2\left(\frac{r}{k}\right)}{r^4}+\frac{k^2 \log \left(\frac{r}{k}\right)}{r^4}+\frac{k l^2 \log \left(\frac{r}{k}\right)}{r^3}+\frac{k l \log \left(\frac{r}{k}\right)}{r^3}+\frac{k m^2 \log \left(\frac{r}{k}\right)}{r}-\frac{2 k M}{r^4}+\frac{4 k M \log \left(\frac{r}{k}\right)}{r^4}+\frac{k}{r^3}-\frac{k \log \left(\frac{r}{k}\right)}{r^3}=
     \frac{k \left(\log \left(\frac{r}{k}\right) \left(-k \log \left(\frac{r}{k}\right)+k+\left(l^2+l-1\right) r+m^2 r^3+4 M\right)-2 M+r\right)}{r^4}\,.
\end{equation}
In Fig. \ref{PotentialS} we plot their behaviour. For k=0.5 (left panel), the event horizon is located at $r_h=1.463$.
Note that near the horizon the blue line  representing to $V_{Sch}$ is negative until $r=2M$, and the red line  representing to $V_{dM}$ is positive. Beyond $r=2M$, both potentials are positive ($V_{Sch}$, and $V_{dM}$).  Then, for $k=2.0$ (central panel), and $r > rh$  with $r_h=2.0$, both potentials are positive ($V_{Sch}$, and $V_{dM}$). Then, for $k=5.0$ (right panel), and $r > rh$  with $r_h=3.618$), beyond the horizon the blue line ($V_{Sch}$) is positive, and the red line ($V_{dM}$) is negative. Beyond $V_{dM}=0$, both potentials are positive ($V_{Sch}$, $V_{dM}$). It is worth mentioning that for large values of $r$, $V_{dM}$ goes to zero, and $V_{eff}(r)$  goes to $V_{Sch}$.

As we can see from the above figure at fixed value of the scalar field mass $m=1$, at large distances the PFDM does not contribute and we recover the Schwarzschild black hole while near the horizon the $k$ parameter of the PFDM plays a decisive role.
%As the  $k$ parameter increases the horizon is also increases as we have shown in Fig. \ref{F9}.
Varying the parameter $k$, from $k=0.5$ to $k=5$ there is an interplay of the potentials from negative to positive values indicating how the PFDM from small to large values interacts with the Schwarzschild black hole.

%Mejorar redaccion As the  $k$ parameter increases the horizon is also increases as we have shown in Fig. \ref{F9}. eso no siempre es asi ... salvo para los k de la fig ... QUIZAS QUEDA MEJOR COMO: \textbf{When the PFDM parameter meets the condition $k > k_{h}$, the event horizon will enlarge.} ... lo borre mejor ...

\begin{figure}[h]
\begin{center}
\includegraphics[width=0.3\textwidth]{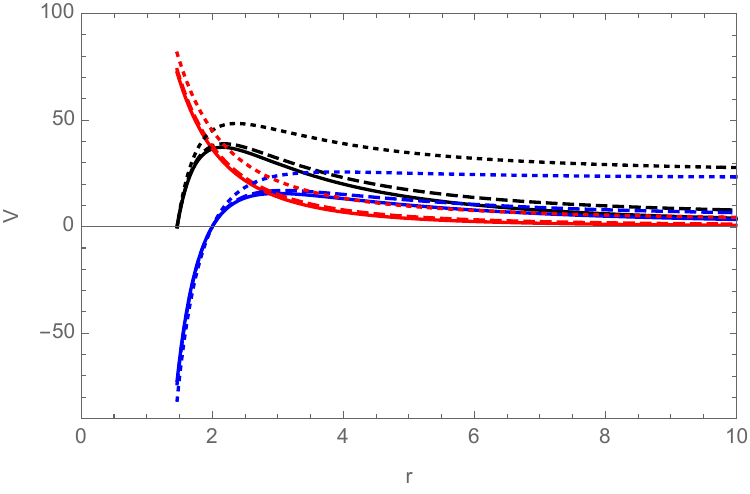}
\includegraphics[width=0.3\textwidth]{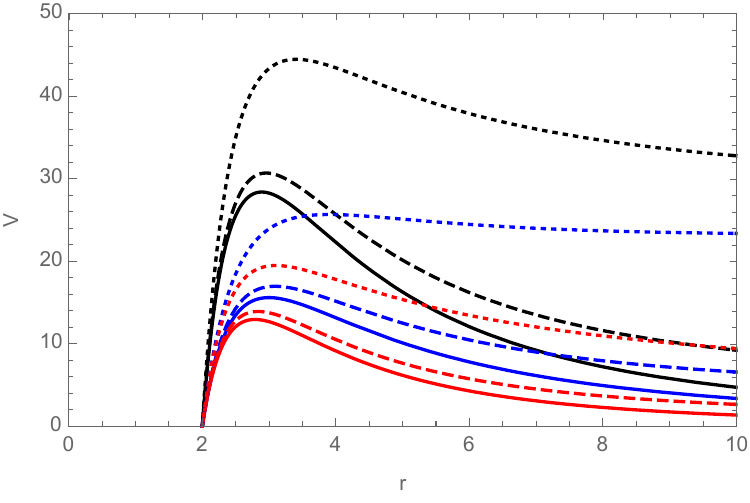}
\includegraphics[width=0.3\textwidth]{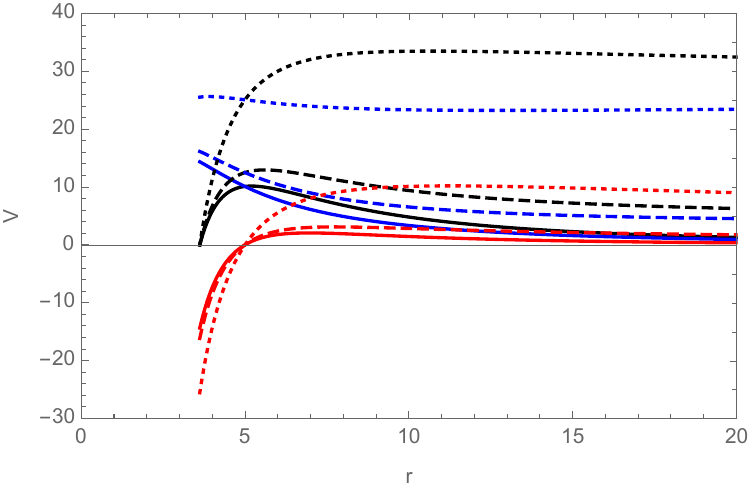}
\end{center}
\caption{The behaviour of the potentials $V_{Sch}$, $V_{dM}$ and $V=V_{Sch}+V_{dM}$ as a function of $r$ with $M=1$, and $\ell=20$. Black line for $V$, blue line for $V_{Sch}$, and red line for $V_{dM}$. Left panel for $k=0.5$, and $r_h=1.463$, central panel for $k=2.0$, and $r_h=2$, and right panel for $k=5.0$, and $r_h=3.618$. Continuous line for massless scalar field $m=0$, dashed line for $m=2.0$, and dotted line for $m=5.0$.}
\label{Potentialmass}
\end{figure}

%\newpage
In Fig. \ref{Potentialmass} we show the effect of the scalar field mass on the potential for the three cases previously analyzed ($k=0.5,2.0,5.0$). We observe that the absolute value of each potential, $V_{eff}(r)$, $V_{Sch}$ and $V_{dM}$ increases, when the scalar field mass increases and this is happening near the horizon of the black hole. We also observe that the largest increase of the absolute value of each potential occurs at intermediate values of the PFDM parameters $k$.  Also, when $r\rightarrow \infty$ we find from Eqs. (\ref{pot}), (\ref{VSch}), and (\ref{VdM}), that the asymptotic behaviour of the potentials are given by $V_{eff}(r)\rightarrow m^2$, $V_{Sch}\rightarrow m^2$ and $V_{dM} \rightarrow 0$ as expected.

\section{Photon sphere modes}
\label{WKBJ}

In this section, in order to get some analytical insight of the behaviour of the QNFs in the eikonal limit $\ell \rightarrow \infty$, we use the method based on the Wentzel-Kramers-Brillouin (WKB) approximation initiated by Mashhoon \cite{Mashhoon} and by Schutz and Iyer \cite{Schutz:1985km}. Iyer and Will computed the third order correction \cite{Iyer:1986np}, and then it was
%Konoplya
extended to the sixth order \cite{Konoplya:2003ii}, and recently up to the 13th order \cite{Matyjasek:2017psv}, see also \cite{Konoplya:2019hlu}.
%has been done by Matyjasek and Opala [12] (see also [13]).

This method has been used to determine the QNFs for asymptotically flat and asymptotically de Sitter black holes. This is due to the WKB method can be used for effective potentials which have the form of potential barriers that approach to a constant value at the horizon and spatial infinity \cite{Konoplya:2011qq}.
However,
only the photon sphere modes can be obtained with this method. The QNMs are determined by the behaviour of the effective potential near its maximum value $V(r^*_{max})$. The Taylor series expansion of the potential around its maximum is given by
\begin{equation}
V(r^*)= V(r^*_{max})+ \sum_{i=2}^{\infty} \frac{V^{(i)}}{i!} (r^*-r^*_{max})^{i} \,,
\end{equation}
where
\begin{equation}
V^{(i)}= \frac{d^{i}}{d r^{*i}}V(r^*)|_{r^*=r^*_{max}}\,,
\end{equation}
corresponds to the $i$-th derivative of the potential with respect to $r^*$ evaluated at the position of the maximum of the potential $r^*_{max}$. Using the WKB approximation up to third order beyond the eikonal limit, the QNFs are given by the following expression \cite{Hatsuda:2019eoj}

\begin{eqnarray}
\omega^2 &=& V(r^*_{max})  -2 i U \,,
\end{eqnarray}
where
\begin{eqnarray}
\notag U &=&  N\sqrt{-V^{(2)}/2}+\frac{i}{64} \left( -\frac{1}{9}\frac{V^{(3)2}}{V^{(2)2}} (7+60N^2)+\frac{V^{(4)}}{V^{(2)}}(1+4 N^2) \right) +\frac{N}{2^{3/2} 288} \Bigg( \frac{5}{24} \frac{V^{(3)4}}{(-V^{(2)})^{9/2}} (77+188N^2)  \\
\notag && +\frac{3}{4} \frac{V^{(3)2} V^{(4)}}{(-V^{(2)})^{7/2}}(51+100N^2) +\frac{1}{8} \frac{V^{(4)2}}{(-V^{(2)})^{5/2}}(67+68 N^2)+\frac{V^{(3)}V^{(5)}}{(-V^{(2)})^{5/2}}(19+28N^2)+\frac{V^{(6)}}{(-V^{(2)})^{3/2}} (5+4N^2)  \Bigg)\,,
\end{eqnarray}
and $N=n+1/2$, with $n=0,1,2,\dots$, is the overtone number.
The imaginary and real part of the QNFs can be written as
\begin{eqnarray}
\label{im} \omega_I^2 &=& - (Im(U)+V/2)+\sqrt{(Im(U)+V/2)^2+Re(U)^2} \,, \\
\omega_R^2 &=& -Re(U)^2 / \omega_I^2 \,,
\end{eqnarray}
respectively, where $Re(U)$ is the real part of $U$ and $Im(U)$ its imaginary part.
Now, defining $L^2= \ell (\ell+1)$, we find that for large values of $L$, the maximum of the potential is approximately at
%\begin{equation}
%r_{max} \approx 3M-\frac{M}{3 L^2} (1-27 m^2 M^2 +18 \Lambda M^2) \mathcal{B} \,,
%\end{equation}
\begin{equation}
r_{max} \approx r_{0}+\frac{1}{L^2}r_{1}\,,
\end{equation}
where
\begin{equation}
    r_{0} =\frac{3}{2} k W\left(\frac{2}{3}e^{\frac{2M}{k}+\frac{1}{3}}\right)\,,
\end{equation}
%\begin{equation}
%r_{0}=\frac{3}{2} k W\left(\frac{2 %e^{\frac{1}{k}+\frac{1}{3}}}{3 k}\right)\,,
%\end{equation}
$W(x)$ is the Lambert function, and
\begin{equation}
 r_{1}=\frac{2r_{0}(k^{2}+12 kM+16 M^{2}-2(2k+3M)r_{0}+m^{2}(k+2M)r_{0}^{3})+2k r_{0}\log(\frac{r_{0}}{k})(-6k-16M+3r_{0}-m^{2}r_{0}^{3}+4k\log(\frac{r_{0}}{k}))}{2r_{0}(2k-6M+4r_{0}+3k\log(\frac{r_{0}}{k}))}\,.
\end{equation}

%\begin{equation}
%  r_{1}=\frac{-\left((k+1) m^2 r_{0}^3\right)+k \log (r_{0}) \left(-4 k \log (r_{0})+6 k+m^2 r_{0}^3-3 r_{0}+8\right)+4 k r_{0}-k (k+6)+3 r_{0}-4}{12 k \log (r_{0})-7 k+6 (r_{0}-2)}  
%\end{equation}

%\begin{equation} \label{coa}
%V(r^*_{max}) \approx \mathcal{B}\left( \frac{L^2}{27 M^2}+\frac{2+ 27 m^2 M^2 -18 \Lambda M^2}{81 M^2} \right) \,,
%\end{equation}
So, the potential evaluated at  $r^*_{max}$, is given by
\begin{eqnarray}
 V(r^*_{max})\approx \frac{(-2M+r_{0}+k\log(\frac{r_{0}}{k}))}{r_{0}^{3}}L^{2}+\frac{1}{9r_{0}^{4}}\Biggl(-3(29 k^{2}+66 kM+60M^{2})+4(-7k+6M)r_{0}-14r_{0}^{2}\nonumber \\
 -18m^{2}(k+2M)r_{0}^{3}+3m^{2}r_{0}^{4}+3k\log(\frac{r_{0}}{k})(33k+60M-4r_{0}+6m^{2}r_{0}^{3}-15k\log(\frac{r_{0}}{k}))\nonumber \\
+\frac{(3k+2r_{0})(61 k^{2}+4r_{0}^{2}(7+3m^{2}r_{0}^{2})+kr_{0}(82+15m^{2}r_{0}^{2}))}{2k+6M-4r_{0}+3k\log(\frac{r_{0}}{k})}\Biggr) \,.
\end{eqnarray}

%\begin{eqnarray}
% V(r^*_{max})\approx \frac{L^2 (k \log (r_{0})+r_{0}-1)}{r_{0}^3}+\frac{-(5k+16) m^2 r_{0}^4+(9-(k-3)k)m^2 r_{0}^3+
% k \log (r_{0}) \left(-3(k+6) m^2 r_{0}^3-k \log (r_{0}) \left(3k-9m^2 r_{0}^3+r_{0}\right)-2(k-1) r_{0}+2 k (k+3)+16 m^2 r_{0}^4\right)-2 k r_{0}^2-(k-1)^2 r_{0}-k (k(k+2)+3)+6 m^2 r_{0}^5}{r_{0}^4(12 k \log (r_{0})-7k+6 (r_{0}-2))}
%\end{eqnarray}

\begin{figure}[h]
\begin{center}
\includegraphics[width=0.4\textwidth]{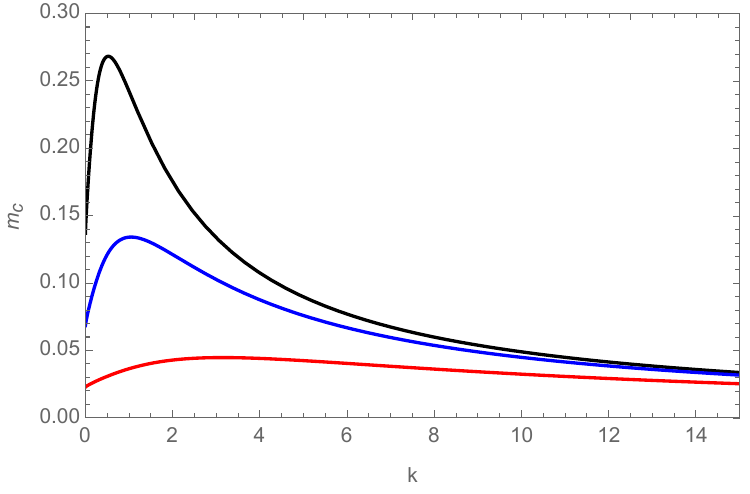}
\end{center}
\caption{The behaviour of the critical scalar field mass $m_{c}$ for the fundamental mode $n=0$ as a function of $k$. Black line for $M=0.5$, blue line for $M=1.0$, and red line for $M=3.0$.}
\label{criticalmass}
\end{figure}

The higher order derivatives  $V^{(i)}(r*_{max})$ with $i=2,3,4,5,6$ are  presented in appendix \ref{HDP}. So, by using these terms, and Eq. (\ref{omegawkb}),  we can find an analytical expression for the critical scalar field mass $m_c$, which is too lengthy to be presented here, because it is obtained from the term proportional to $1/L^2$ in $\omega$.  However, we plot its behaviour as a function of $k$ in Fig. \ref{criticalmass}. We can observe that when the BH mass $M$ increases the critical scalar field mass decreases. Also, there is a value of $k_c$ corresponding to a maximal value for the critical scalar field mass. Namely, as the PFDM parameter $k$ increases in the interval $k < k_c$, $m_c$ increases and reach their maximum values at $k = k_c$. Then, for $k > k_c$ the $m_c$  decreases when the parameter $k$ increases. Remarkably, there is a value of the PFDM parameter $k$, for which the critical scalar field mass is the same that for the  Schwarzschild BH, beyond this value of $k$ the critical scalar field mass is smaller than of the Schwarzschild background.  This behaviour of  the scalar field mass $m$ and the PFDM parameter $k$  indicates that these is  an interplay between the real matter and the phantom matter effecting the mass of the background Schwarzschild black hole.
%\newpage

\begin{figure}[h]
\begin{center}
\includegraphics[width=0.3\textwidth]{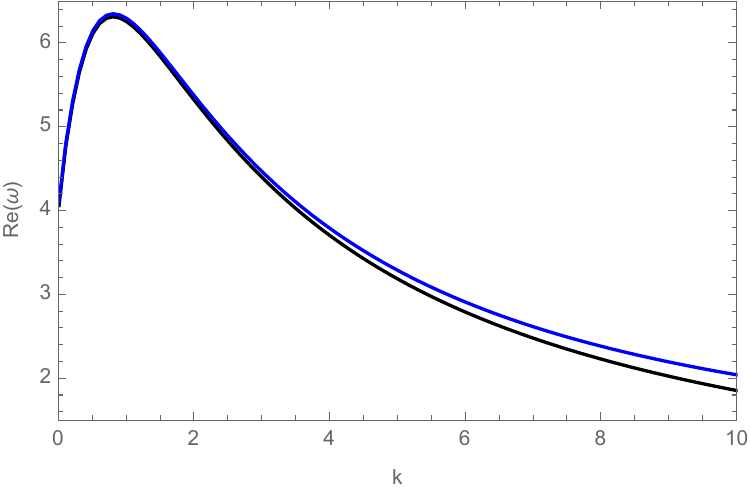}
\includegraphics[width=0.3\textwidth]{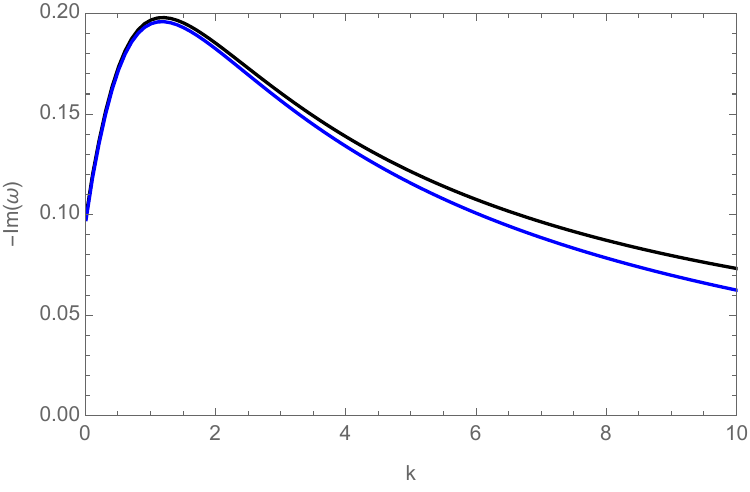}
\end{center}
\caption{The behaviour of $Re(\omega)$ (left panel), and $Im(\omega)$ (right panel) for the fundamental mode ($n=0$) as a function of the PFDM intensity parameter $k$ with $M=1$, and $\ell=20$. Black line for massless  scalar field ($m=0$), and blue line for massive scalar field ($m=1.0$).}
\label{ReIm}
\end{figure}

Now, we plot in Fig.~\ref{ReIm} the behaviour of the real and imaginary part of the QNFs as a function of the PFDM intensity parameter $k$, separately. We can observe  that
%there is
%a reflecting point $k_0$ corresponding to maximal values for the real part of the QNF.
there is a value of the parameter \textbf{$k_0\approx 0.81$}, for which the real part of the QNFs is maximum, which is associated with a reflecting point in the context of BH shadow. Namely, as the PFDM parameter $k$ increases in the interval $k < k_0$, the QNF increases and reach their maximum values at $k = k_0$, as was point out in Ref. \cite{Jusufi:2019ltj} for low values of $\ell$. Then, for $k > k_0$ the real part of the QNF decreases when the parameter $k$ increases. Remarkably, there is a value of the PFDM parameter $k$, for which the real part of the QNF is the same that for the  Schwarzschild BH, beyond this value of $k$ the frequency of oscillation is smaller than the Schwarzschild case.  Also, a similar behaviour occurs for the absolute value of the imaginary part of the QNF. Also, it is possible to observe that for massive scalar field, the real part of the QNFs increases and the absolute value of the imaginary part of the QNFs decreases in comparison with massless scalar field.

\newpage

We show in the following tables that the real part of  the QNFs  have a maximum at $k_0 \approx 0.81$, and  this value does not depend on $n$ (see Table \ref{Table1}) or $\ell$  (see Table \ref{Table3}). Note  also that the imaginary part of  the QNFs  have a maximum at $k \approx 1.19$, and  this value  also does not depend on $n$ (see Table \ref{Table2}) or $\ell$  (see Table \ref{Table4}).

\begin{table}[h]
\caption{QNFs for $M =1$, $\ell=20$, $m=0$ and different values of $k$, and $n$.
}
\label{Table1}\centering
%\scalebox{1} {
%\resizebox{\columnwidth}{!}{
\begin{tabular}{ | c | c | c | c | }
\hline
$k$ & $n=0$ & $n=1$ & $n=2$ \\ \hline
  $0.80$ & $6.312337229-0.1915424746 i$ & $6.305063321-0.5749004929 i$ & $6.290555033-0.9590774523 i$    \\
 $0.81$&  $6.312403646-0.1919177167 i$& $6.305095686-0.5760279567 i$ & $6.290519578-0.9609623546 i$     \\
   $0.82$ & $6.312128105-0.192279291 i$ & $6.304786728-0.577114394 i$ & $6.290144067-0.9627788027 i$ \\\hline
\end{tabular}
\end{table}

\begin{table}[h]
\caption{The fundamental QNFs for $M =1$, $m=0$ and different values of $k$, and $\ell$.
}
\label{Table3}\centering
%\resizebox{\columnwidth}{!}{
\begin{tabular}{ | c | c | c | c | }
\hline
$k$ & $\ell=20$ & $\ell=40$ & $\ell=60$ \\ \hline
  $0.80$ & $6.312337229-0.1915424746 i$ & $12.46960361-0.191520156 i$ & $18.62711699-0.1915159121 i$    \\
  $0.81$ & $6.312403646-0.1919177167 i$ & $12.46973043-0.1918952382 i$ & $18.6273052-0.1918909639 i$    \\
   $0.82$ & $6.312128105-0.192279291 i$ & $12.46918178-0.1922566546 i$ & $18.62648438-0.1922523502 i$    \\\hline
\end{tabular}
\end{table}

\begin{table}[h]
\caption{QNFs for $M =1$, $\ell=20$, $m=0$ and different values of $k$, and $n$.
}
\label{Table2}\centering
%\resizebox{\columnwidth}{!}{
\begin{tabular}{ | c | c | c | c | }
\hline
$k$ & $n=0$ & $n=1$ & $n=2$ \\ \hline
  $1.18$ & $6.137647315-0.1980102076 i$ & $6.129473798-0.5943544216 i$ & $6.11317502-0.991669935 i$    \\
  $1.19$ & $6.129636142-0.1980113209 i$ & $6.121448324-0.5943587095 i$ & $6.105121121-0.9916802457 i$     \\
   $1.20$ & $6.121513845-0.1980059472 i$ & $6.113312125-0.5943435188 i$ & $6.096957289-0.9916580336 i$    \\\hline
\end{tabular}
%}
\end{table}

\begin{table}[h]
\caption{The fundamental QNFs for $M =1$, $m=0$ and different values of $k$, and $\ell$.
}
\label{Table4}\centering
%\resizebox{\columnwidth}{!}{
\begin{tabular}{ | c | c | c | c | }
\hline
$k$ & $\ell=20$ & $\ell=40$ & $\ell=60$ \\ \hline
  $1.18$ & $6.137647315-0.1980102076 i$ & $12.12437267-0.1979831253 i$ & $18.1113699-0.1979779755 i$   \\
  $1.19$ & $6.129636142-0.1980113209 i$ & $12.10854406-0.1979841471 i$ & $18.08772422-0.1979789798 i$ \\
  $1.20$ & $6.121513845-0.1980059472 i$ & $12.09249597-0.1979786833 i$ & $18.06375069-0.197973499 i$    \\\hline
\end{tabular}
%}
\end{table}

\newpage

\begin{figure}[h]
\begin{center}
\includegraphics[width=0.3\textwidth]{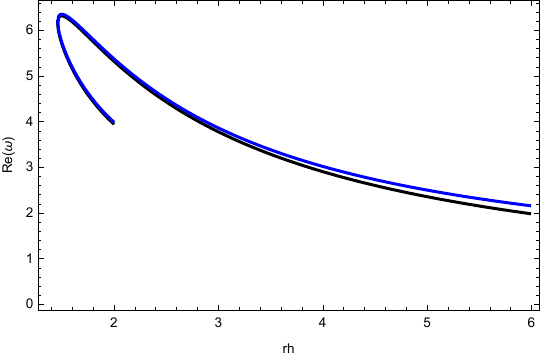}
\includegraphics[width=0.3\textwidth]{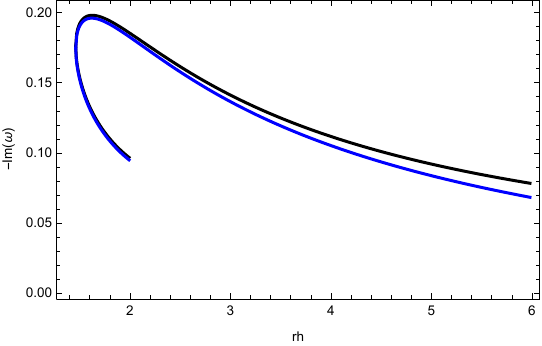}
\end{center}
\caption{The behaviour of $Re(\omega)$ (left panel), and $Im(\omega)$ (right panel) for the fundamental mode ($n=0$) as a function of the event horizon $r_h$  with $M=1$, and $\ell=20$. Black line for massless  scalar field ($m=0$), and blue line for massive scalar field ($m=1.0$).}
\label{ReImrh}
\end{figure}

Now, we show the  relation of QNFs  to the location of the horizon in Fig. \ref{ReImrh}. We can observe a maximum value of the real part of the QNFs at $r_h \approx 1.502$, and for the imaginary part at $r_h \approx 1.622$. Also, it is possible to observe that for massive scalar fields, the real part of the QNFs increases and the absolute value of the imaginary part of the QNFs decreases in comparison  to massless scalar fields.

%\newpage

\begin{figure}[h]
\begin{center}
\includegraphics[width=0.32\textwidth]{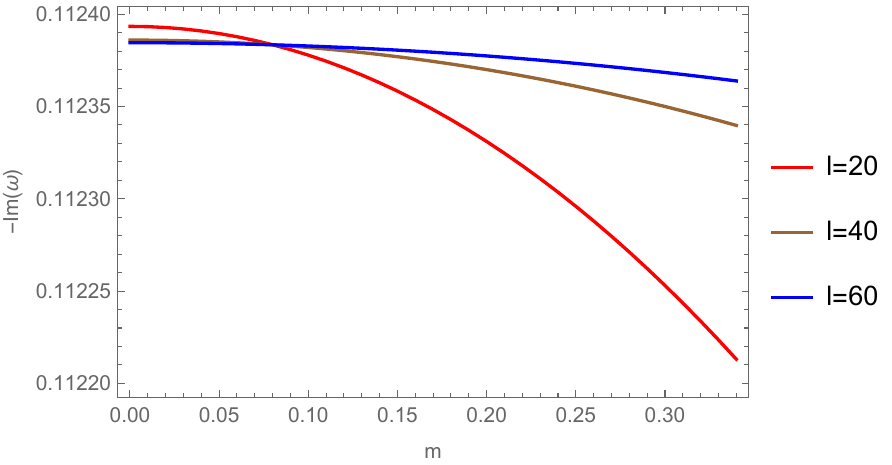}
\includegraphics[width=0.32\textwidth]{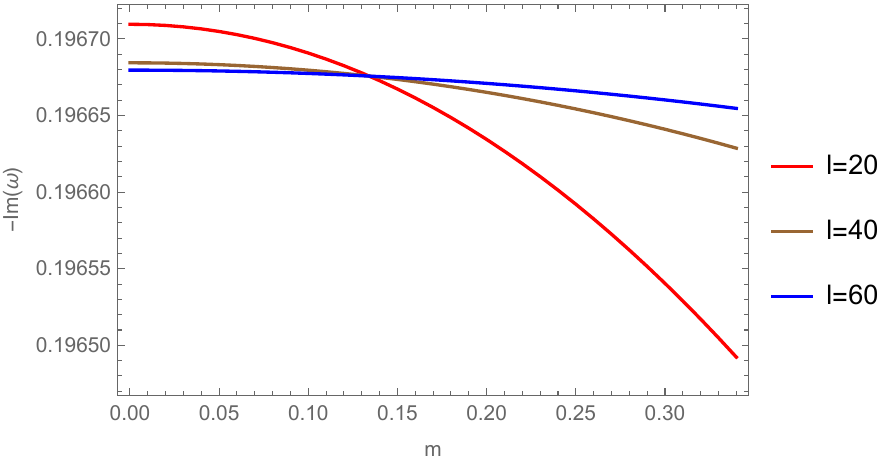}
\includegraphics[width=0.32\textwidth]{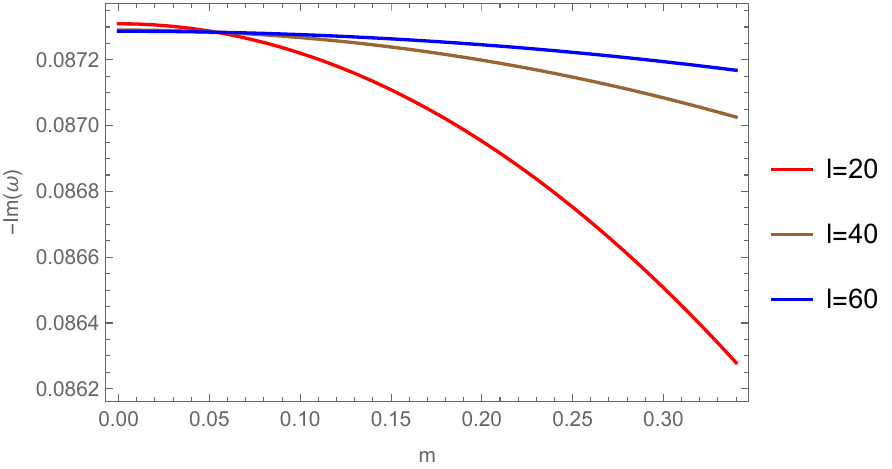}
\end{center}
\caption{The behaviour of $-Im(\omega)$ for the fundamental mode ($n=0$) as a function of the scalar field mass $m$ for different values of the parameter $\ell=20,40,60$, with $M=1$, $k=0.07$ (left panel), $k=1.0$ (central panel), and  $k=8.0$ (right panel). Here, the WKB method gives $m_{c}\approx 0.08$, $m_{c}\approx 0.13$, and $m_{c}\approx 0.05$, respectively.}
\label{AB1}
\end{figure}

\begin{figure}[h]
\begin{center}
\includegraphics[width=0.32\textwidth]{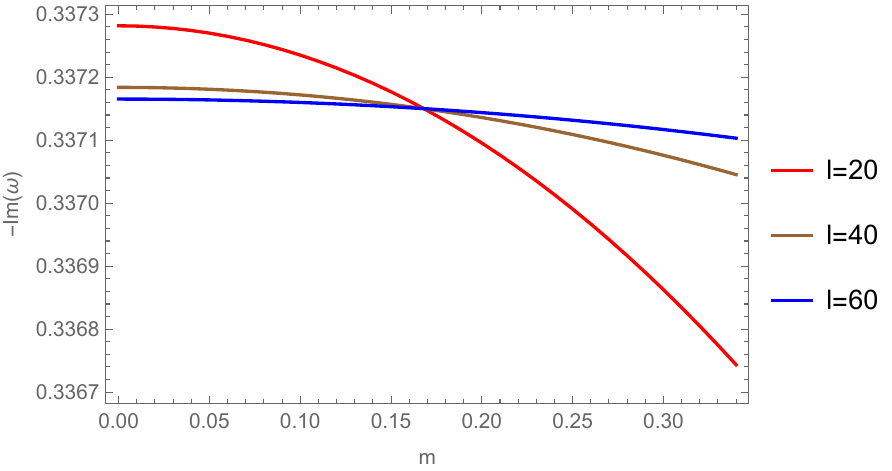}
\includegraphics[width=0.32\textwidth]{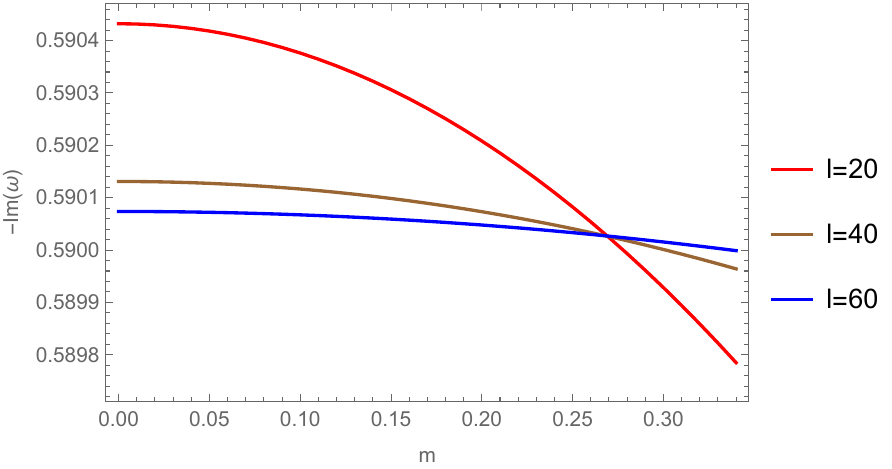}
\includegraphics[width=0.32\textwidth]{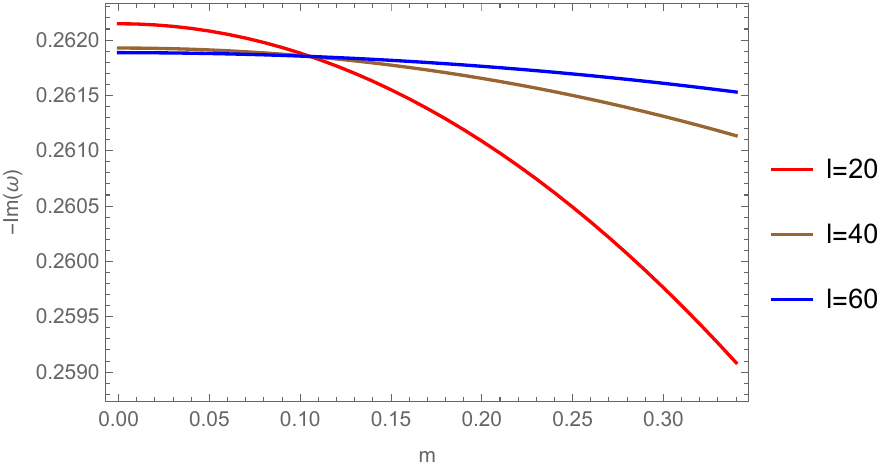}
\end{center}
\caption{The behaviour of $-Im(\omega)$ for the first overtone ($n=1$)  as a function of the scalar field mass $m$ for different values of the parameter $\ell=20,40,60$, with $M=1$, $k=0.07$ (left panel), $k=1.0$ (central panel), and  $k=8.0$ (right panel). Here, the WKB method gives $m_{c}\approx 0.17$, $m_{c}\approx 0.27$, and $m_{c}\approx 0.10$, respectively.}
\label{AB2}
\end{figure}

Now, in order to show the anomalous behaviour, we plot in Figs. \ref{AB1}, and \ref{AB2}, the behaviour of $-Im(\omega)$ as a function of $m$ by using the 6th order WKB  method for $n=0$, and $n=1$, respectively. We can observe an anomalous decay rate, i.e,  for $m<m_c$, the longest-lived modes are the ones with highest angular number $\ell$; whereas, for $m>m_c$, the longest-lived modes are the ones with smallest angular number. Also, when the overtone number $n$ increases the parameter $m_c$ increases. Also, it is possible to observe that the behaviour of the critical scalar field mass with respect to the parameter $k$  agrees with Fig. \ref{criticalmass}.

\newpage

Finally, in order to analyze the behaviour of the longest-lived modes with respect to the scalar field mass and the PFDM parameter $k$, we show in Tables \ref{Tablekm1} and \ref{Tablekm2} some fundamental QNFs for $M=1$, and different values of $m$, and $k$.  Firstly, it is possible to observe the anomalous behaviour of the decay rate of the QNFs previously described. Bold letters show the longest-lived modes. Additionally, a particular behaviour can be observed for fixed values of the scalar field mass, see $m=0.08$, and $m=0.09$. In these cases, for small values of the PFDM parameter $k$ (see $k=0.01,0.07$), the longest-lived modes are the one with smallest angular number, then as the $k$ parameter increases (see $k=1,2$) the longest-lived modes are the one with highest angular number, and then for biggest values of $k$ (see $k=5,10$), the longest-lived modes are the one with smallest angular number.

\begin{table}[ht]
\caption{Fundamental QNFs for $M =1$, and different values of $k$ and $m$. Bold letters indicate the longest-lived modes.
}
\label{Tablekm1}\centering
\resizebox{\columnwidth}{!}{
\begin{tabular}{ | c | c | c | c | c | c | c | c |  }
\hline
$k$ & $\ell$ & $m=0$ & $m=0.07$ & $m=0.08$ & $m=0.09$ & $m=0.10$ & $m=0.11$  \\ \hline
\multirow{3}{0.55cm}{$0.01$} & $20$ & 4.060774422-0.099298049 $i$ & 4.060975957-0.099289838 $i$ & \bf{4.061037652-0.099287324 $i$} & \bf{4.061107574-0.099284475 $i$} & \bf{4.061185722-0.099281291 $i$} & \bf{4.061272097-0.099277772 $i$} \\
 & $40$ & 8.022042946-0.099291809 $i$ & 8.022145063-0.099289704 $i$ & 8.022176324-0.099289060 $i$ &  8.022211752-0.099288329 $i$&  8.022251349-0.099287513 $i$& 8.022295114-0.099286611 $i$ \\
 & $60$ & \bf{11.983414279-0.099290623 $i$} & \bf{11.983482651-0.099289679 $i$} & 11.983503582-0.099289391 $i$ & 11.983527303-0.099289063 $i$ & 11.983553815-0.099288697 $i$ & 11.983583118-0.099288293 $i$
\\\hline
 \multirow{3}{0.55cm}{$0.07$} & $20$ & 4.519531072-0.112393405 $i$ & 4.519716287-0.112385764 $i$ & \bf{4.519772985-0.112383426 $i$} & \bf{4.519837244-0.112380775 $i$}  & \bf{4.519909064-0.112377812 $i$} & \bf{4.519988443-0.1123745378 $i$} \\
 & $40$ & 8.928296798-0.112385968 $i$ & 8.928390648-0.112384009 $i$ & 8.928419377-0.112383409 $i$ & 8.928451938-0.112382730 $i$ &8.928488329-0.112381970 $i$ & 8.928528551-0.112381131 $i$ \\
 & $60$ & \bf{13.337180946-0.112384554 $i$} & \bf{13.337243784-0.112383676 $i$} & 13.337263020-0.112383407 $i$ & 13.337284821-0.112383103 $i$ &13.337309187-0.112382762 $i$ & 13.337336118-0.112382386 $i$
\\\hline
\multirow{3}{0.2cm}{$1$} & $20$ & 6.258068954-0.196709555 $i$ & 6.258256572-0.196700346 $i$ & 6.258314006-0.196697527 $i$ & 6.258379098-0.196694332 $i$  & 6.258451849-0.196690761 $i$   & 6.258532257-0.196686814 $i$  \\
 & $40$ & 12.362318873-0.196684403 $i$ & 12.362413986-0.196682041
 $i$ &12.362443102-0.196681318 $i$ & 12.362476100-0.196680499 $i$ & 12.362512981-0.196679583 $i$ &  12.362553744-0.196678571 $i$ \\
 & $60$ & \bf{18.466831759-0.196679621 $i$} & \bf{18.466895449-0.196678562 $i$} &\bf{18.466914945-0.196678238 $i$} & \bf{18.466937042-0.196677871 $i$} & \bf{18.466961737-0.196677460 $i$} & \bf{18.466989033-0.196677006 $i$} \\ \hline
\multirow{5}{0.2cm}{$2$} & $20$ &5.321892057-0.185273143 $i$ & 5.322151017-0.185259388 $i$ &5.322230291-0.185255177 $i$ & 5.322320135-0.185250405 $i$ &  5.322420549-0.185245072 $i$ & 5.322531533-0.185239177 $i$  \\
 & $40$ & 10.512743590-0.185242445 $i$ & 10.512874906-0.185238916 $i$ & 10.512915105-0.185237836 $i$ & 10.512960663-0.185236612 $i$ & 10.513011582-0.185235244 $i$ & 10.513067860-0.185233732 $i$ \\
 & $60$ & \bf{15.703870347-0.185236608 $i$} & \bf{15.703958283-0.185235026 $i$} &  \bf{15.703985202-0.185234542 $i$} & \bf{15.704015710-0.185233993 $i$} & \bf{15.704049808-0.185233380 $i$} & \bf{15.704087494-0.185232702 $i$}   \\ \hline
 \multirow{5}{0.2cm}{$5$} & $20$ &3.186561101-0.121455537 $i$ & 3.187062385-0.121426919 $i$ & \bf{3.187215840-0.121418159 $i$} & \bf{3.187389757-0.121408231 $i$} & \bf{ 3.187584136-0.121397135 $i$} & \bf{3.187798976-0.121384871 $i$}  \\
 & $40$ & 6.294508804-0.121430476 $i$ & 6.294763071-0.121423133 $i$ & 6.294840908-0.121420885 $i$ &6.294929123-0.121418337 $i$ & 6.295027717-0.121415490 $i$ & 6.295136689-0.121412343 $i$ \\
 & $60$ & \bf{9.402654865-0.121425710 $i$} & \bf{9.402825145-0.121422418 $i$} & 9.402877271-0.121421411 $i$ & 9.402936348-0.121420269 $i$ & 9.403002375-0.121418992 $i$ & 9.403075352-0.121417582 $i$ \\ \hline
 \multirow{5}{0.2cm}{$10$} & $20$ &1.853977904-0.073298439 i & \bf{1.854893653-0.073244586 $i$} & \bf{1.855173990-0.073228102 $i$} & \bf{1.855491709-0.073209422 $i$} & \bf{1.855846810-0.073188545 $i$} & \bf{1.856239296-0.073165473 $i$}  \\
 & $40$ & 3.662177134-0.073281951 $i$ & 3.662641688-0.073268131 $i$ &  3.662783899-0.073263900 $i$ &3.662945073-0.073259106 $i$ & 3.663125208-0.073253747 $i$ &3.663324305-0.073247825 $i$ \\
 & $60$ & \bf{5.470500775-0.073278815 $i$} & 5.470811889-0.073272620 $i$ & 5.470907128-0.073270724 $i$ & 5.471015066-0.073268575 $i$ & 5.471135703-0.073266173 $i$ & 5.471269038-0.073263518 $i$ \\\hline
\end{tabular}
}
\end{table}

\begin{table}[ht]
\caption{Fundamental QNFs for $M =1$, and different values of $k$ and $m$. Bold letters indicate the longest-lived modes.
}
\label{Tablekm2}\centering

\scalebox{0.7} {

%\resizebox{\columnwidth}{!}{
\begin{tabular}{ | c | c | c | c | c | c |c |   }
\hline
$k$ & $\ell$ & $m=0.12$ & $m=0.13$ & $m=0.14$  & $m=0.15$  & $m=0.30$ \\ \hline
\multirow{3}{0.55cm}{$0.01$} & $20$ & \bf{4.061366698-0.099273917 $i$} & \bf{4.061469527-0.099269727 $i$} & \bf{4.061580583-0.099265202 $i$} & \bf{4.061699867-0.099260342 $i$}  & \bf{4.064476621-0.099147190 $i$} \\
 & $40$ &8.022343047-0.099285622 $i$ & 8.022395148-0.099284548 $i$ & 8.022451417-0.099283388 $i$ & 8.022511855-0.099282142 $i$ & 8.023918636-0.099253137 $i$ \\
 & $60$ &11.983615211-0.099287850 $i$ &11.983650096-0.099287369 $i$ & 11.983687771-0.099286849 $i$ & 11.983728237-0.099286290 $i$ & 11.984670127-0.099273291 $i$
\\\hline
 \multirow{3}{0.55cm}{$0.07$} & $20$& \bf{4.520075383-0.112370951 $i$} & \bf{4.520169884-0.112367053 $i$} & \bf{4.520271945-0.112362843 $i$} & \bf{4.520381567-0.112358321 $i$}  & \bf{4.522933357-0.112253048 $i$} \\
 & $40$ & 8.928572603-0.112380211 $i$ & 8.928620486-0.112379212 $i$ & 8.928672200-0.112378132 $i$ & 8.928727745-0.112376973 $i$ & 8.930020627-0.112349986 $i$ \\
 & $60$ & 13.337365613-0.112381974 $i$ &13.337397674-0.112381526 $i$ & 13.337432299-0.112381042 $i$ & 13.337469489-0.112380522 $i$ & 13.338335129-0.112368427 $i$
\\\hline
\multirow{3}{0.2cm}{$1$} & $20$ &6.258620324-0.196682492 $i$ &6.258716049-0.196677793 $i$ & \bf{6.258819432-0.196672719 $i$} & \bf{6.258930473-0.196667269 $i$}  & \bf{6.261515146-0.196540410 $i$} \\
 & $40$ &12.362598389-0.196677462 $i$ &12.362646916-0.196676257 $i$ & 12.362699325-0.196674955 $i$ & 12.362755617-0.196673557 $i$ & 12.364065865-0.196641019 $i$ \\
 & $60$ &\bf{8.467018928-0.196676509 $i$} &\bf{18.467051422-0.196675969 $i$} & 18.467086517-0.196675386 $i$ & 18.467124210-0.196674759 $i$ & 18.468001568-0.196660175 $i$
\\\hline
\multirow{5}{0.2cm}{$2$}  & $20$ &5.322653087-0.185232721 $i$ & \bf{5.322785212-0.185225703 $i$} & \bf{5.322927907-0.185218124 $i$} & \bf{5.323081173-0.185209984 $i$}  & \bf{5.326648698-0.185020522 $i$} \\
 & $40$ &10.513129498-0.185232075 $i$ &10.513196496-0.185230275 $i$ & 10.513268854-0.185228331 $i$ & 10.513346572-0.185226242 $i$ & 10.515155538-0.185177635 $i$ \\
 & $60$ &\bf{15.704128770-0.185231959 $i$} &15.704173636-0.185231153 $i$ & 15.704222090-0.185230281 $i$ & 15.704274134-0.185229345 $i$ & 15.705485500-0.185207557 $i$
\\\hline
 \multirow{5}{0.2cm}{$5$} & $20$ &\bf{3.188034279-0.121371440 $i$} & \bf{3.188290044-0.121356841 $i$} & \bf{3.188566272-0.121341075 $i$} & \bf{3.188862964-0.121324141 $i$}  & \bf{3.195769311-0.120930137 $i$} \\
 & $40$ &6.295256040-0.121408896 $i$ &6.295385769-0.121405150 $i$ & 6.295525877-0.121401104 $i$ & 6.295676363-0.121396758 $i$ & 6.299179135-0.121295618 $i$ \\
 & $60$ &9.403155279-0.1214160367 $i$ &9.403242157-0.121414357 $i$ & 9.403335985-0.121412544 $i$ & 9.403436763-0.121410596 $i$ &  9.405782484-0.121365256 $i$
\\\hline
 \multirow{5}{0.2cm}{$10$} & $20$ & \bf{1.856669167-0.073140205 $i$} & \bf{1.857136425-0.073112743 $i$} & \bf{1.857641073-0.073083086 $i$} & \bf{1.858183112-0.073051236 $i$}  & \bf{1.870802383-0.072310891 $i$} \\
 & $40$ &3.663542365-0.073241339 $i$ &3.663779387-0.073234288 $i$ & 3.664035373-0.073226674 $i$ & 3.664310321-0.073218497 $i$  & 3.670710327-0.073028218 $i$\\
 & $60$ &5.471415072-0.073260610 $i$ &5.471573805-0.073257449 $i$ & 5.471745237-0.073254036 $i$ & 5.471929367-0.073250370 $i$ & 5.476215275-0.073165050 $i$
\\\hline
\end{tabular}
}
\end{table}

%\clearpage

\newpage
\section{Conclusions}
\label{conclusion}

In this work, we studied the propagation of massive scalar fields  in the background of BHs immersed in perfect fluid dark matter through the QNFs by using the WKB method in order to determine if there is an anomalous decay behaviour  in the QNMs as it was observed in other BH backgrounds. Here, we considered the photon sphere modes in our analysis, which are complex.

Concerning to the photon sphere modes, we showed that
%there is
%a reflecting point $k_0$ corresponding to maximal values for the real part of the QNF.
there is a value of the parameter $k=k_0$, for which the real part of the QNFs is maximum, which is associated with a reflecting point in the context of BH shadow. Namely, as the PFDM parameter $k$ increases in the interval $k < k_0$, the QNF increases and reach their maximum values at $k = k_0$. Then, for $k > k_0$ the real part of the QNF decreases when the parameter $k$ increases. Remarkably, there is a value of the PFDM parameter $k$, for which the real part of the QNF is the same that for the  Schwarzschild BH, beyond this value of $k$ the frequency of oscillation is smaller than the Schwarzschild case.  Also, a similar behaviour occurs for the absolute value of the imaginary part of the QNF. However, there is not a value of $k$, such that will be possible to recover the  QNFs for the Schwarzschild background, i.e the real and imaginary part of the QNFs. Also, for massive scalar field, the real part of the QNFs increases and the absolute value of the imaginary part of the QNFs decreases in comparison with massless scalar field.

We showed the existence of anomalous decay rate of QNMs, i.e, the absolute values of the imaginary part of the QNFs decay when the angular harmonic numbers increase if the mass of the scalar field is smaller than a critical mass. On the contrary  they grow when the angular harmonic numbers increase, if the mass of the scalar field is larger than the critical mass and they also increase with the overtone number $n$, for $\ell \geq n$. Also, the critical scalar field mass decreases  when the BH mass $M$ increases, and there is a value of $k_c$ corresponding to a maximal value for the critical scalar field mass. Namely, as the PFDM parameter $k$ increases in the interval $k < k_c$, the $m_c$ increases and reach their maximum values at $k = k_c$. Then, for $k > k_c$ the $m_c$  decreases when the parameter $k$ increases. Remarkably, there is a value of the PFDM parameter $k$, for which the critical scalar field mass is the same that for the  Schwarzschild BH, beyond this value of $k$ the critical scalar field mass is smaller than for the Schwarzschild background.

Also, we have reported a particular behaviour for fixed values of the scalar field mass ($m=0.08, 0.09$), where  for small values of the PFDM parameter $k$ ($k=0.01,0.07$), the longest-lived modes are the ones with smallest angular number, then as the $k$ parameter increases ($k=1,2$) the longest-lived modes are the ones with highest angular number, and then for biggest values of $k$ ($k=5,10$), the longest-lived modes are the ones with smallest angular number. Also, in order to show that the scale of the location of the horizon is not important due to the behaviour for $\bar{k}<1$ is similar to the behaviour  for $\bar{k}>1$, with $\bar{k}=k/r_h$, we have included a dimensionless analysis, see appendix \ref{DA}.

It would be interesting to consider charged black holes immersed in perfect fluid dark matter %\cite{Xu:2016ylr,Shaymatov:2020wtj}, 
that represent to the RN dS BHs when the PFDM parameter $k\rightarrow 0$, in order to analyze if it is possible to avoid the existence of unstable modes for $\ell=0$ for a value of the parameter $k$ for charged massive scalar field. Also, to study the superradiance, as well as, the existence of  bound states which could to trigger an instability, work in this direction is in progress.

\appendix{}

\section{Higher order derivatives of the potential}
\label{HDP}
In order to explicitly set forth the terms of the
%sixth order
WKB approximation outlined in Section IV, we present the higher order derivatives  $V^{(i)}(r*_{max})$ with $i=2,3,4,5,6$ of the potential.
\begin{eqnarray}
\notag V^{(0)} &=& V_{02} L^2 + V_{00} + \mathcal{O}(L^{-2})\,,\\
\notag V^{(2)} &=& V_{22}L^2 + V_{20} + \mathcal{O}(L^{-2})\,, \\
\notag V^{(3)} &=& V_{32}L^2 + \mathcal{O}(L^0)\,,\\
\notag V^{(4)} &=& V_{42}L^2+ \mathcal{O}(L^0)\,,\\
\notag V^{(5)} &=& V_{52}L^2 + \mathcal{O}(L^0)\,,\\
V^{(6)} &=& V_{62}L^2 +\mathcal{O}(L^0)\,,
\end{eqnarray}
where
\begin{equation}
    V_{00}=\frac{\eta  \left(-\eta +k+m^2 r_0^3+4 M-3 r_1-r_0\right)-\left(2 M-r_0\right) \left(k+m^2 r_0^3+2 M\right)+r_1 \left(k+6 M-2 r_0\right)}{r_0^4}\,,
\end{equation}
\begin{equation}
V_{02} =\frac{-2M+r_{0}+\eta}{r_{0}^{3}}\,,\\
\end{equation}

\begin{equation}
  V_{22} =  \frac{\left(\eta-2 M+r_0\right) \left(k^2+ \eta \left(15  \eta -11 k+20 (r_0-3 M)\right)+22 k M-9 k r_0+60 M^2-40 M r_0+6 r_0^2\right)}{r_0^7}\,,
\end{equation}

\begin{eqnarray}
 \nonumber V_{20} & = & \frac{1}{r_0^9}\Bigg(r_0 r_1 \left(-8 \eta +k+16 M-7 r_0\right) \left(k^2+\eta  \left(15 \eta -11 k+20 \left(r_0-3 M\right)\right)+22 k M-9 k r_0+60 M^2-40 M r_0+6 r_0^2\right)\\
 \nonumber && -r_0 \left(-\eta +2 M-r_0\right) \Bigg(k^3+\eta  \Bigg(-18 k^2+\eta  \left(-24 \eta +48 k+3 m^2 r_0^3+144 M-35 r_0+15 r_1\right)\\
 \nonumber && +k \left(-5 m^2 r_0^3-192 M+64 r_0+19 r_1\right)-4 M \left(3 m^2 r_0^3-35 r_0+15 r_1\right)+2 r_0 \left(m^2 r_0^3-6 r_0+20 r_1\right)\\
 \nonumber && -288 M^2\Bigg)+k^2 \left(m^2 r_0^3+36 M-15 r_0-10 r_1\right)+k \Bigg(2 M \left(5 m^2 r_0^3-64 r_0-19 r_1\right)+r_0 \left(-3 m^2 r_0^3+19 r_0+2 r_1\right)\\
&& +192 M^2\Bigg)+4 M^2 \left(3 m^2 r_0^3-35 r_0+15 r_1\right)-4 M r_0 \left(m^2 r_0^3-6 r_0+20 r_1\right)+192 M^3+18 r_0^2 r_1\Bigg)\Bigg)\,,
\end{eqnarray}

\begin{eqnarray}
  \notag  V_{32}&=&-\frac{1}{{r_0^9}}\left(\eta -2 M+r_0\right) \Bigg(-k^3+\eta  \Bigg(29 k^2+\eta  \left(105 \eta -122 k+210 \left(r_0-3 M\right)\right)+488 k M-190 k r_0+1260 M^2\\
\notag  &&  -840 M r_0+130 r_0^2\Bigg)+k^2 \left(26 r_0-58 M\right)+k \left(-488 M^2+380 M r_0-71 r_0^2\right)-840 M^3+840 M^2 r_0\\
 && -260 M r_0^2+24 r_0^3\Bigg)\,,
\end{eqnarray}

\begin{eqnarray}
 \notag   V_{42}&=&\frac{1}{r_0^{11}}(\eta -2 M+r_0) \Bigg(k^4+134 k^3 M+r_0^2 \left(443 k^2+5014 k M+9520 M^2\right)+2508 k^2 M^2\\
   \notag &&+\eta  \Bigg(-67 k^3+\eta  \Bigg(627 k^2+3 \eta  \left(315 \eta -506 k+840 \left(r_0-3 M\right)\right)+k \left(9108 M-3441 r_0\right)\\
\notag &&    +140 \left(162 M^2-108 M r_0+17 r_0^2\right)\Bigg)+76 k^2 \left(14 r_0-33 M\right)+k \left(-18216 M^2+13764 M r_0-2507 r_0^2\right)\\
\notag &&+28 \left(-1080 M^3+1080 M^2 r_0-340 M r_0^2+33 r_0^3\right)\Bigg)-r_0 \left(63 k^3+2128 k^2 M+13764 k M^2+20160 M^3\right)\\
&&+12144 k M^3-4 r_0^3 (145 k+462 M)+15120 M^4+120 r_0^4\Bigg)\,,
\end{eqnarray}
\begin{eqnarray}
  \notag  V_{52}&=&-\frac{1}{r_0^{13}}\left(\eta -2 M+r_0\right) \Bigg(-k^5-290 k^4 M-10472 k^3 M^2-103752 k^2 M^3+r_0^3 \left(6631 k^2+62608 k M+105728 M^2\right)\\
 \notag  && -2 r_0^2 \left(1037 k^3+25459 k^2 M+136464 k M^2+176400 M^3\right)+\eta  \Bigg(145 k^4+10472 k^3 M\\
\notag && +r_0^2 \left(25459 k^2+272928 k M+529200 M^2\right)+155628 k^2 M^2+\eta  \Bigg(-2618 k^3+3 \eta  \Bigg(4323 k^2+\eta  \Bigg(3465 \eta -7141 k\\
\notag && -11550 \left(3 M-r_0\right)\Bigg)+k \left(57128 M-21150 r_0\right)+2100 \left(66 M^2-44 M r_0+7 r_0^2\right)\Bigg)+k^2 \left(31787 r_0-77814 M\right)\\
\notag && -12 k \left(42846 M^2-31725 M r_0+5686 r_0^2\right)-56 \left(14850 M^3-14850 M^2 r_0+4725 M r_0^2-472 r_0^3\right)\Bigg)\\
\notag && -2 r_0 \left(2341 k^3+63574 k^2 M+380700 k M^2+554400 M^3\right)+685536 k M^3-56 r_0^3 (559 k+1888 M)+831600 M^4\\
\notag &&+7308 r_0^4\Bigg)+4 r_0 \left(35 k^4+2341 k^3 M+31787 k^2 M^2+126900 k M^3+138600 M^4\right)-342768 k M^4\\
&& -232 r_0^4 (22 k+63 M)-332640 M^5+720 r_0^5\Bigg)\,,
\end{eqnarray}

\begin{eqnarray}
 \notag   V_{62}&=&\frac{1}{r_0^{15}}\left(\eta -2 M+r_0\right) \Bigg(k^6+606 k^5 M+38956 k^4 M^2+687280 k^3 M^3+4411392 k^2 M^4\\
\notag && +r_0^4 \left(96087 k^2+790304 k M+1214640 M^2\right)-2 r_0^3 \left(26415 k^3+519972 k^2 M+2413912 k M^2+2821280 M^3\right)\\
\notag && +8 r_0^2 \left(1037 k^4+47202 k^3 M+512376 k^2 M^2+1763502 k M^3+1732500 M^4\right)\\
\notag && +\eta  \Bigg(-303 k^5+4 k^4 \left(4505 r_0-9739 M\right)-24 k^3 \left(42955 M^2-36978 M r_0+7867 r_0^2\right)\\
\notag && -12 k^2 \left(735232 M^3-876564 M^2 r_0+341584 M r_0^2-43331 r_0^3\right)+\eta  \Bigg(9739 k^4+12 k^3 \left(42955 M-18489 r_0\right)\\
\notag && +24 k^2 \left(275712 M^2-219141 M r_0+42698 r_0^2\right)+\eta  \Bigg(-85910 k^3+3 \eta  \Bigg(91904 k^2+\eta  \Bigg(45045 \eta -113623 k\\
\notag && -180180 \left(3 M-r_0\right)\Bigg)+k \left(1136230 M-414567 r_0\right)+11550 \left(234 M^2-156 M r_0+25 r_0^2\right)\Bigg)\\
\notag && +k^2 \left(876564 r_0-2205696 M\right)-6 k \left(2272460 M^2-1658268 M r_0+293917 r_0^2\right)\\
\notag && -3080 \left(7020 M^3-7020 M^2 r_0+2250 M r_0^2-229 r_0^3\right)\Bigg)+4 k \Bigg(6817380 M^3-7462206 M^2 r_0+2645253 M r_0^2 \\
\notag && -301739 r_0^3\Bigg)+420 \left(77220 M^4-102960 M^3 r_0+49500 M^2 r_0^2-10076 M r_0^3+723 r_0^4\right)\Bigg)\\
\notag && -8 k \left(3408690 M^4-4974804 M^3 r_0+2645253 M^2 r_0^2-603478 M r_0^3+49394 r_0^4\right)\\
\notag && -48 \left(540540 M^5-900900 M^4 r_0+577500 M^3 r_0^2-176330 M^2 r_0^3+25305 M r_0^4-1338 r_0^5\right)\Bigg)\\
\notag && -r_0 \left(297 k^5+36040 k^4 M+887472 k^3 M^2+7012512 k^2 M^3+19899216 k M^4+17297280 M^5\right)+10907808 k M^5\\
&&-4 r_0^5 (12215 k+32112 M)+8648640 M^6+5040 r_0^6\Bigg)\,,
\end{eqnarray}
with
\begin{equation}
    \eta=k\log\left(\frac{r_{0}}{k}\right)\,.
\end{equation}
On the other hand, our interest is to evaluate the QNFs for large values of $L$, so we expand the frequencies as a power series in $L$. It is important to keep in mind that in the eikonal limit, the leading term is linear in $L$, and for $k=0$, we should recover the Schwarzschild black hole frequencies. Next, we consider the following expression in powers of L
\begin{equation}
\label{omegawkb}
  \omega=\omega_{1m}L+\omega_{0}+\omega_{1}L^{-1}+\omega_{2}L^{-2} + \mathcal{O}(L^{-3})\,,
\end{equation}
where
%\begin{equation}
    %\omega_{1m}=\sqrt{\frac{-2M+r_{0}+k\log(\frac{r_{0}}{k})}{r_{0}^{3}}}
%\end{equation}

\begin{equation}
    \omega_{1m}=\frac{1}{r_0}\sqrt{\frac{-2M+r_{0}+\eta}{r_{0}}}\,,
\end{equation}

\begin{equation}
    \omega_0=-\frac{i\sqrt{2} }{4 r_0^2 }\sqrt{-k^2-\eta  \left(15 \eta -11 k+20 \left(r_0-3 M\right)\right)-22 k M+9 k r_0-60 M^2+40 M r_0-6 r_0^2}\,,
\end{equation}
\begin{eqnarray}
    \omega_1&=&\frac{1}{38880 r_0^{5/2} \sqrt{\eta -2 M+r_0}}\Bigg(-52110 \eta ^2+6 \Bigg(-645 k^2+3240 r_1 \left(k+6 M-2 r_0\right)-13905 k M+5959 k r_0\\
    \notag && -6480 m^2 M r_0^3+3240 m^2 r_0^4-34740 M^2+21000 M r_0-2241 r_0^2\Bigg)
     +\frac{1}{\chi}\Bigg(9 k^3 (5110 M-541 k)\\
\notag && +6 k^2 r_0 (12347 M-5736 k)-3 \eta  \left(7665 k^3+12347 k^2 r_0+5757 k r_0^2+700 r_0^3\right)+14 r_0^3 (300 M-1453 k)\\
\notag && +2 k r_0^2 (17271 M-23194 k)-2420 r_0^4\Bigg) +\frac{1}{\chi^2}\Bigg(11 \Bigg(549 k^5 (k+22 M) +3 k^4 r_0 (13808 M-1213 k)\\
\notag && +2 k^3 r_0^2 (27635 M-8257 k)+k^2 r_0^3 (35628 M-23939 k)\\
\notag && -\eta  \left(33 k^2+47 k r_0+16 r_0^2\right) \left(183 k^3+367 k^2 r_0+226 k r_0^2+40 r_0^3\right)+40 r_0^5 (32 M-129 k)+k r_0^4 (10992 M-16145 k)\\
\notag && -624 r_0^6\Bigg)\Bigg)+45 \eta  \left(927 k+8 \left(54 m^2 r_0^3+579 M-175 r_0-162 r_1\right)\right)\Bigg)\,,
\end{eqnarray}
with
\begin{equation}
    \chi=k^2+\eta  (15 \eta -11 k+20 (r_0-3 M))+22 k M-9 k r_0+60 M^2-40 M r_0+6 r_0^2\,.
\end{equation}
\begin{eqnarray}
 \notag   \omega_2 &=& \mu^{-1}i \Bigg(216 r_0 \delta ^6-1728 \Bigg(\left(m^2 r_0^3-r_0+k+4 M-\eta -3 r_1\right) \eta -\left(2 M-r_0\right) \left(m^2 r_0^3+k+2 M\right)\\
\notag && +\left(k+6 M-2 r_0\right) r_1\Bigg) r_0 \delta ^5-108 r_0 \alpha  \delta ^4+1728 \Bigg(r_0 r_1 \left(k+16 M-8 \eta -7 r_0\right) \delta\\
\notag && -r_0 \left(2 M-\eta -r_0\right) \Bigg(k^3+\left(m^2 r_0^3-15 r_0+36 M-10 r_1\right) k^2+\Bigg(192 M^2+2 \left(5 m^2 r_0^3-64 r_0-19 r_1\right) M\\
\notag && +r_0 \left(-3 m^2 r_0^3+19 r_0+2 r_1\right)\Bigg) k+192 M^3+\eta  \Bigg(-18 k^2+\left(-5 m^2 r_0^3+64 r_0-192 M+19 r_1\right) k\\
\notag && -288 M^2+\eta  \left(3 m^2 r_0^3-35 r_0+48 k+144 M-24 \eta +15 r_1\right)-4 M \left(3 m^2 r_0^3-35 r_0+15 r_1\right)\\
\notag && +2 r_0 \left(m^2 r_0^3-6 r_0+20 r_1\right)\Bigg)+18 r_0^2 r_1+4 M^2 \left(3 m^2 r_0^3-35 r_0+15 r_1\right)-4 M r_0 \left(m^2 r_0^3-6 r_0+20 r_1\right)\Bigg)\Bigg) \delta ^4\\
\notag && +132 \sigma ^2 r_0 \delta ^3+36 \Bigg(k^6+606 M k^5+38956 M^2 k^4+687280 M^3 k^3+4411392 M^4 k^2+10907808 M^5 k+8648640 M^6\\
\notag && +5040 r_0^6-4 (12215 k+32112 M) r_0^5+\left(96087 k^2+790304 M k+1214640 M^2\right) r_0^4\\
\notag && -2 \left(26415 k^3+519972 M k^2+2413912 M^2 k+2821280 M^3\right) r_0^3+8 \Bigg(1037 k^4+47202 M k^3+512376 M^2 k^2\\
&& +1763502 M^3 k+1732500 M^4\Bigg) r_0^2+\eta  \Bigg(-303 k^5+4 \left(4505 r_0-9739 M\right) k^4-24 \Bigg(42955 M^2-36978 r_0 M\\
\notag && +7867 r_0^2\Bigg) k^3-12 \left(735232 M^3-876564 r_0 M^2+341584 r_0^2 M-43331 r_0^3\right) k^2-8 \Bigg(3408690 M^4-4974804 r_0 M^3\\
\notag && +2645253 r_0^2 M^2-603478 r_0^3 M+49394 r_0^4\Bigg) k+\eta  \Bigg(9739 k^4+12 \left(42955 M-18489 r_0\right) k^3\\
\notag && +24 \left(275712 M^2-219141 r_0 M+42698 r_0^2\right) k^2+4 \left(6817380 M^3-7462206 r_0 M^2+2645253 r_0^2 M-301739 r_0^3\right) k\\
\notag && +\eta  \Bigg(-85910 k^3+\left(876564 r_0-2205696 M\right) k^2-6 \left(2272460 M^2+293917 r_0^2-1658268 Mr_0\right) k\\
\notag && +3 \eta  \Bigg(91904 k^2+\left(1136230 M-414567 r_0\right) k+\eta  \left(-113623 k+45045 \eta -180180 \left(3 M-r_0\right)\right)\\
\notag && +11550 \left(234 M^2-156 r_0 M+25 r_0^2\right)\Bigg)-3080 \left(7020 M^3-7020 r_0 M^2+2250 r_0^2 M-229 r_0^3\right)\Bigg)\\
\notag && +420 \left(77220 M^4-102960 r_0 M^3+49500 r_0^2 M^2-10076 r_0^3 M+723 r_0^4\right)\Bigg)-48 \Bigg(540540 M^5-900900 r_0 M^4\\
\notag && +577500 r_0^2 M^3-176330 r_0^3 M^2+25305 r_0^4 M-1338 r_0^5\Bigg)\Bigg)-\Bigg(297 k^5+36040 M k^4+887472 M^2 k^3\\
\notag && +7012512 M^3 k^2+19899216 M^4 k+17297280 M^5\Bigg) r_0\Bigg) r_0 \delta ^3-63 r_0 \alpha ^2 \delta ^2-156 r_0 \sigma  \Bigg(-k^5-290 M k^4-10472 M^2 k^3\\
\notag && -103752 M^3 k^2-342768 M^4 k-332640 M^5+720 r_0^5-232 (22 k+63 M) r_0^4+\left(6631 k^2+62608 M k+105728 M^2\right) r_0^3\\
\notag && -2 \left(1037 k^3+25459 M k^2+136464 M^2 k+176400 M^3\right) r_0^2+\eta  \Bigg(145 k^4+10472 M k^3+155628 M^2 k^2+685536 M^3 k\\
\notag && +831600 M^4+7308 r_0^4-56 (559 k+1888 M) r_0^3+\left(25459 k^2+272928 M k+529200 M^2\right) r_0^2\\
\notag && +\eta  \Bigg(-2618 k^3+\left(31787 r_0-77814 M\right) k^2-12 \left(42846 M^2-31725 r_0 M+5686 r_0^2\right) k\\
\notag && +3 \eta  \left(4323 k^2+\left(57128 M-21150 r_0\right) k+\eta  \left(-7141 k+3465 \eta -11550 \left(3 M-r_0\right)\right)+2100 \left(66 M^2-44 r_0 M+7 r_0^2\right)\right)\\
\notag && -56 \left(14850 M^3-14850 r_0 M^2+4725 r_0^2 M-472 r_0^3\right)\Bigg)-2 \left(2341 k^3+63574 M k^2+380700 M^2 k+554400 M^3\right) r_0\Bigg)\\
\notag && +4 \left(35 k^4+2341 M k^3+31787 M^2 k^2+126900 M^3 k+138600 M^4\right) r_0\Bigg) \delta ^2-155 r_0 \sigma ^4+342 \alpha  \delta \sigma ^2 r_0\Bigg)\,,
\end{eqnarray}
with
\begin{equation}
    \mu=
    %=6912 \sqrt{2} r_0^9 \sqrt{\frac{-2 M+\eta +r_0}{r_0^3}} \delta ^4 \sqrt{-\frac{\left(-2 %M+\eta +r_0\right) \delta }{r_0^7}}
    6912 r_0^4 \left(-2 M+\eta +r_0\right) \delta ^4 \sqrt{-2\delta }\,,
\end{equation}
\begin{eqnarray}
   \xi =15 \eta -11 k+20 \left(r_0-3 M\right) \,,
\end{eqnarray}

\begin{eqnarray}
    \delta =\eta  \xi +k^2+22 k M-9 k r_0+60 M^2-40 M r_0+6 r_0^2\,,
\end{eqnarray}

\begin{eqnarray}
\notag    \sigma &=&-k^3+\eta  \left(29 k^2+\eta  \left(105 \eta -122 k+210 \left(r_0-3 M\right)\right)+488 k M-190 k r_0+1260 M^2-840 M r_0+130 r_0^2\right)\\
 &&   +k^2 \left(26 r_0-58 M\right)+k \left(-488 M^2+380 M r_0-71 r_0^2\right)-840 M^3+840 M^2 r_0-260 M r_0^2+24 r_0^3\,,
\end{eqnarray}

\begin{eqnarray}
 \notag   \alpha &=&k^4+134 k^3 M+r_0^2 \left(443 k^2+5014 k M+9520 M^2\right)+2508 k^2 M^2\\
 \notag &&   +\eta  \Bigg(-67 k^3+\eta  \Bigg(627 k^2+3 \eta  \left(315 \eta -506 k+840 \left(r_0-3 M\right)\right)+k \left(9108 M-3441 r_0\right)\\
\notag  && +140 \left(162 M^2-108 M r_0+17 r_0^2\right)\Bigg)+76 k^2 \left(14 r_0-33 M\right)+k \left(-18216 M^2+13764 M r_0-2507 r_0^2\right)\\
\notag && +28 \left(-1080 M^3+1080 M^2 r_0-340 M r_0^2+33 r_0^3\right)\Bigg)-r_0 \left(63 k^3+2128 k^2 M+13764 k M^2+20160 M^3\right)\\
&&+12144 k M^3-4 r_0^3 (145 k+462 M)+15120 M^4+120 r_0^4\,.
\end{eqnarray}

\section{Dimensionless analysis.}
\label{DA}

\begin{figure}[h!]
\begin{center}
\includegraphics[width=0.3\textwidth]{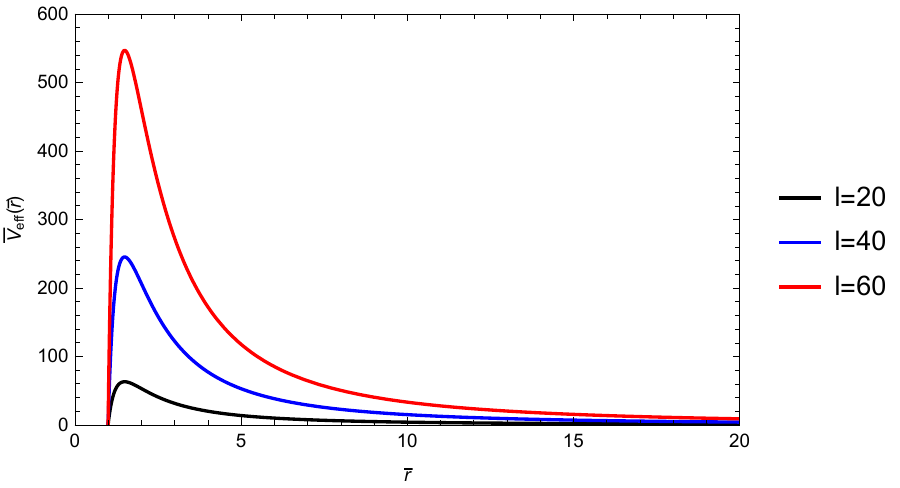}
\includegraphics[width=0.3\textwidth]{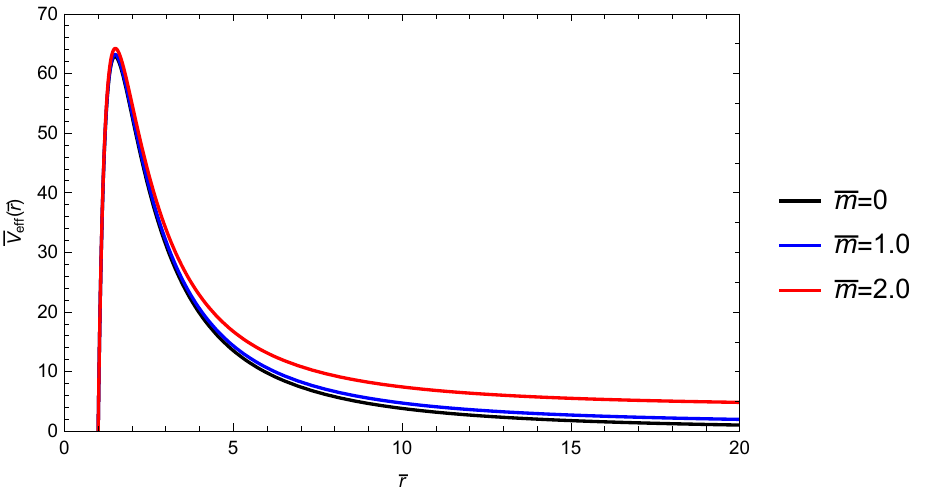}
\includegraphics[width=0.3\textwidth]{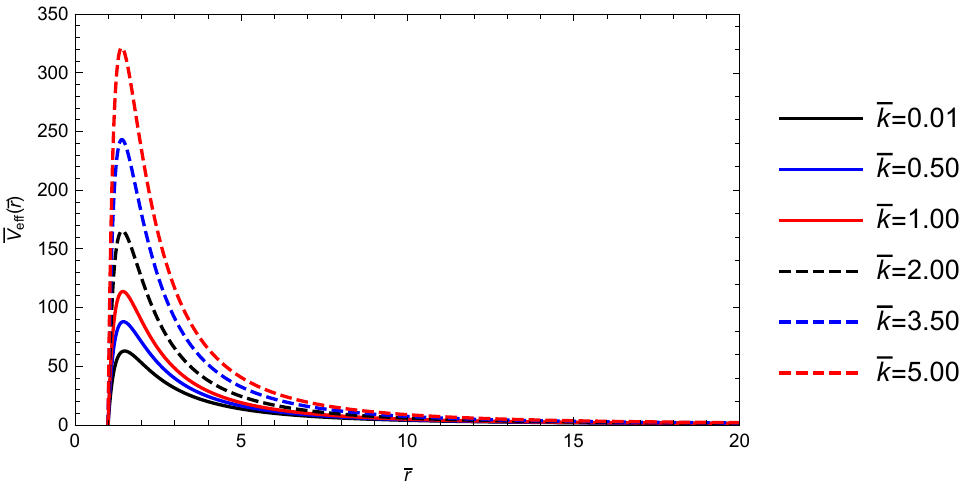}
\end{center}
\caption{The effective potential $\bar{V}_{eff}$ as a function of $\bar{r}$. Left panel for massless scalar field $\bar{m}=0$, $\bar{k}=0.01$, and $\ell=20,40,60$. Central panel for $\ell=20$, $\bar{k}=0.01$, and  $\bar{m}=0,1.0,2.0$. Right panel for $\ell=20$, $\bar{m}=0$, and $\bar{k}=0.01, 0.50, 1.00, 2.00, 3.50, 5.00$.}
\label{Potential2}
\end{figure}

Scaling by the event horizon radius $r_h$, that is, by considering $\bar{r}=r/r_h$, $\bar{k}=k/r_h$, and $\bar{m}=mr_h$. The metric function and the effective potential can be written as
\begin{equation}
f(\bar{r})=1-\frac{1}{\bar{r}}+\frac{\bar{k}}{\bar{r}}\ln(\bar{r})\,,\label{metfun2}
\end{equation}
\begin{equation}
 \bar{V}_{eff}(\bar{r})=\frac{f(\bar{r})}{\bar{r}^2} \left(\kappa^{2} + \bar{m}^2 \bar{r}^2 +\bar{r}f^\prime(\bar{r})\right)=r_h^2 V_{eff}(r)~,
 \end{equation}
 respectively. Here, the prime denotes  a derivative with respect to $\bar{r}$. In Fig. \ref{Potential2}, we show   the  behaviour  of the effective potential for different values of the parameters. Note that a barrier of potential occurs  in each case, with  its maximum value increasing when $\ell$ or $\bar{m}$ increases. However, $\bar{V}_{eff}(\bar{r})$ has a different behaviour
 than ${V}_{eff}({r})$, the maximun value of the barrier of potential increases when $\bar{k}$ increases.

\newpage

Also, it is possible to find the critical  mass parameter  $\bar{m}_c$,  which for small values of $\bar{k}$  is given perturbatively by the expansion

\begin{eqnarray}
    \label{mcbar}
  \bar{m}_c &=& \frac{1}{27}\sqrt{\frac{137}{10}}+ \bar{k} \left(\frac{77}{30 \sqrt{1370}}+\frac{1}{27} \sqrt{\frac{137}{10}} \ln \left(\frac{3}{2}\right)\right)\\
 \nonumber && + \bar{k}^2 \left(-\frac{450343}{277425 \sqrt{1370}}+\frac{1}{27} \sqrt{\frac{137}{10}} \ln ^2\left(\frac{3}{2}\right)+\frac{4}{135} \sqrt{\frac{2}{685}} \ln \left(\frac{3}{2}\right)\right)+O\left(\bar{k}^3\right)\,.
\end{eqnarray}

\begin{figure}[h]
\begin{center}
\includegraphics[width=0.3\textwidth]{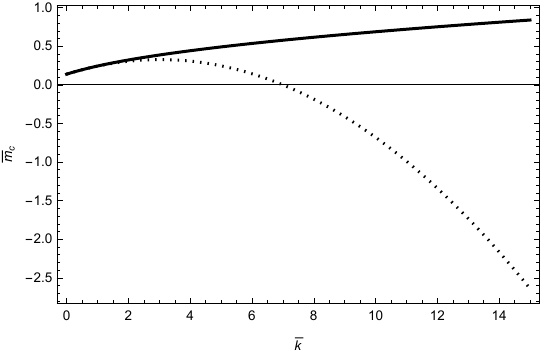}
\end{center}
\caption{The behaviour of the critical scalar field mass $\bar{m}_{c}$ for the fundamental mode $n=0$ as a function of $\bar{k}$.
 Continuous line for the exact value  of $\bar{m}_{c}$, and dotted line  for approximate expression of $\bar{m}_{c}$ via Eq. (\ref{mcbar}).}
\label{criticalmass2}
\end{figure}

We plot its behaviour as a function of $\bar{k}$ in Fig. \ref{criticalmass2}. We can observe that when the parameter $\bar{k}$ increases the critical mass $\bar{m}_c$ increases and also for $\bar{k}=0$ we recover the value of Schwarzschild black hole.

\begin{figure}[h]
\begin{center}
\includegraphics[width=0.3\textwidth]{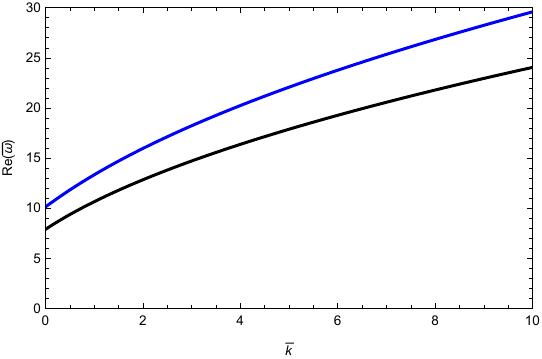}
\includegraphics[width=0.3\textwidth]{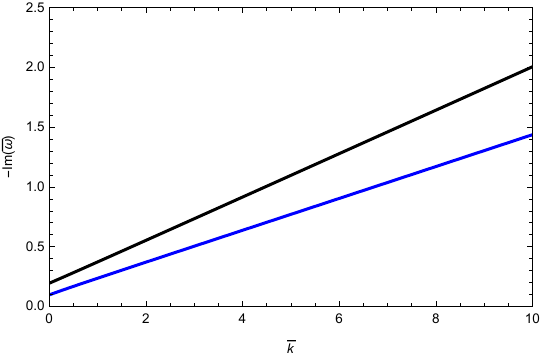}
\end{center}
\caption{The behaviour of $Re(\bar{\omega})$ (left panel), and $Im(\bar{\omega})$ (right panel) for the fundamental mode ($n=0$) as a function of the
%PFDM intensity
parameter $\bar{k}$,
%with $M=1$,
and $\ell=20$. Black line for
%massless  scalar field (
$\bar{m}=0$, and blue line for
%massive scalar field (
$\bar{m}=10$.}
\label{ReIm2}
\end{figure}

Now, we plot in Fig.~\ref{ReIm2} the behaviour of the real and imaginary  parts of $\bar{\omega}=\omega r_h$ as a function of the parameter $\bar{k}$, separately. Here,  both the real part and absolute value of the imaginary part of $\bar{\omega}$ increase when the parameter $\bar{k}$ or $\bar{m}$ increases, i.e. there is a corresponding elevation in both oscillation frequency and damping. Specifically, when $\bar{k} > 1$, it signifies a scenario predominantly governed by dark matter, as this condition is equivalent to $k > r_h$. In this context, the influence of dark matter manifests as an augmentation in both the oscillation frequency and damping of the QNFs. 
% \textbf{It looks like from Fig.~\ref{ReIm2} that for $\bar{k}>1$, there is a higher frequency of oscillation, which corresponds to the scenario dominated by dark matter since $\bar{k}>1$ is equivalent to saying $k>r_{h}$.}

\begin{figure}[h]
\begin{center}
\includegraphics[width=0.32\textwidth]{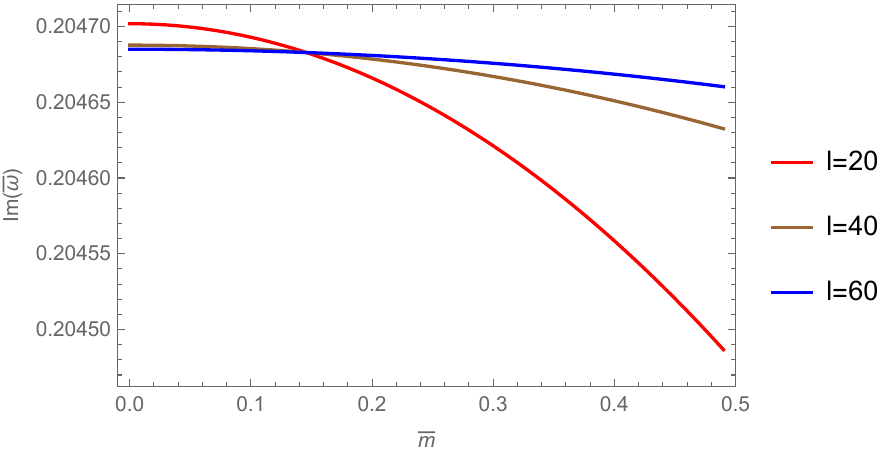}
\includegraphics[width=0.32\textwidth]{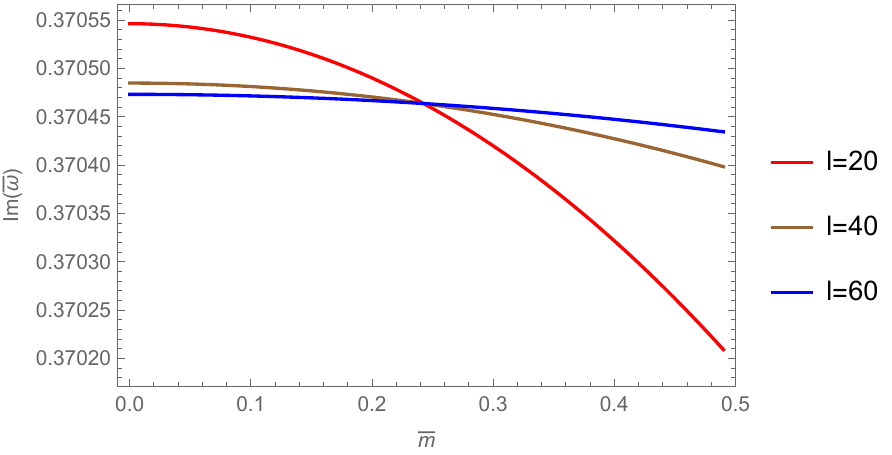}
\includegraphics[width=0.32\textwidth]{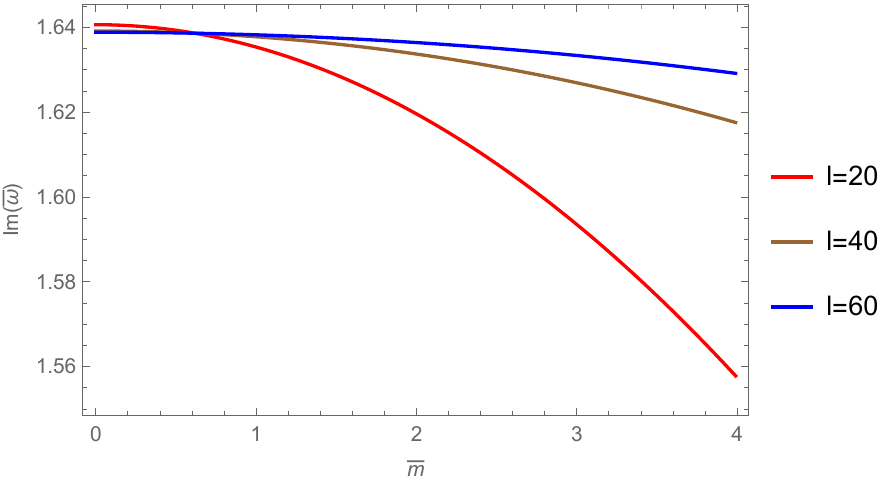}
\end{center}
\caption{The behaviour of $-Im(\bar{\omega})$ for the fundamental mode ($n=0$) as a function of 
%the scalar field mass 
$\bar{m}$ for different values of the parameter $\ell=20,40,60$, with , $\bar{k}=0.07$ (left panel), $\bar{k}=1.0$ (central panel), and  $\bar{k}=8.0$ (right panel). Here, the WKB method gives $\bar{m}_{c}\approx 0.15 $, $\bar{m}_{c}\approx 0.24$, and $\bar{m}_{c}\approx 0.62$, respectively.}
\label{AB12}
\end{figure}

\begin{figure}[h]
\begin{center}
\includegraphics[width=0.32\textwidth]{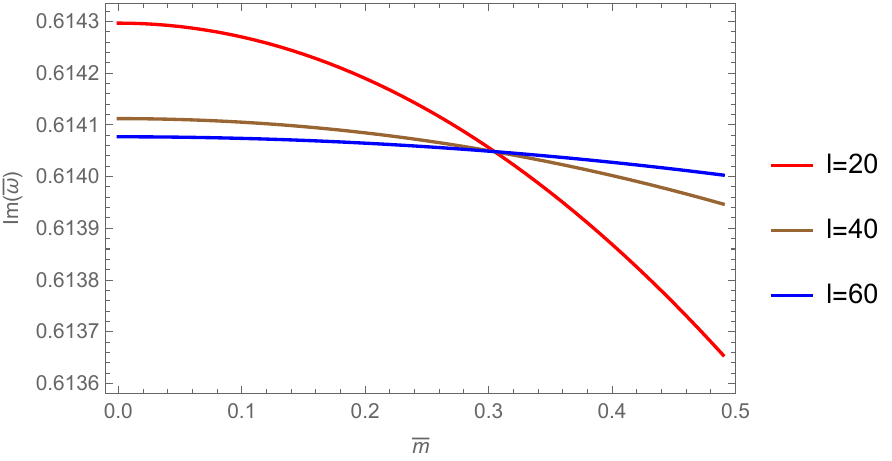}
\includegraphics[width=0.32\textwidth]{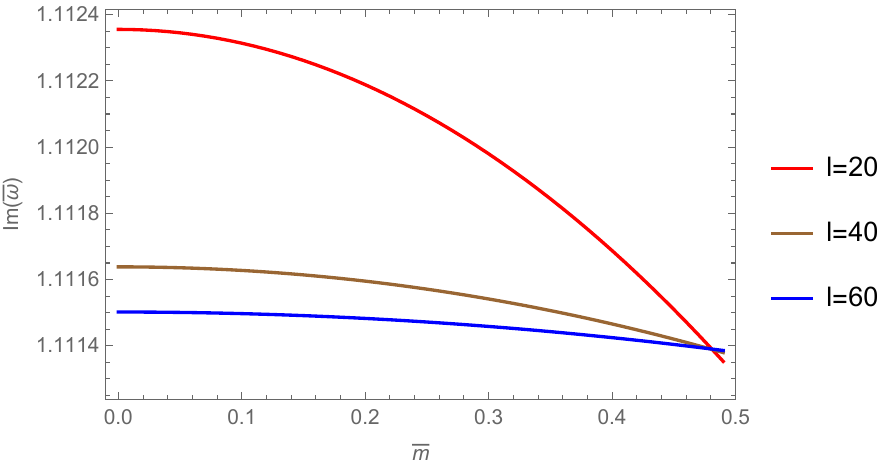}
\includegraphics[width=0.32\textwidth]{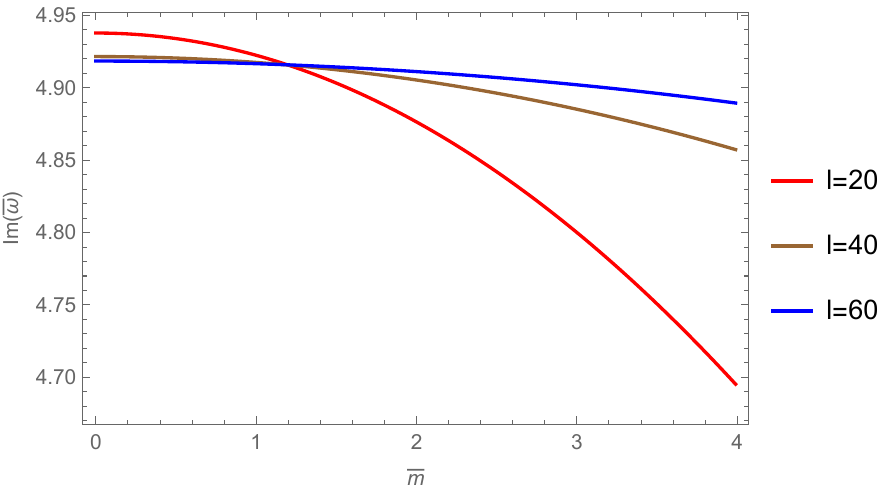}
\end{center}
\caption{The behaviour of $-Im(\bar{\omega})$ for the first overtone ($n=1$)  as a function of 
%the scalar field mass 
$\bar{m}$ for different values of the parameter $\ell=20,40,60$, with $M=1$, $\bar{k}=0.07$ (left panel), $\bar{k}=1.0$ (central panel), and  $\bar{k}=8.0$ (right panel). Here, the WKB method gives $\bar{m}_{c}\approx 0.30 $, $\bar{m}_{c}\approx 0.48 $, and $\bar{m}_{c}\approx 1.19 $, respectively.}
\label{AB22}
\end{figure}

Now, in order to show the anomalous behaviour, we plot in Fig. \ref{AB12}, and \ref{AB22}, the behaviour of $-Im(\bar{\omega})$ as a function of $\bar{m}$ by using the 6th order WKB  method for $n=0$, and $n=1$, respectively. We can observe an anomalous decay rate, i.e,  for $\bar{m}<\bar{m}_c$, the longest-lived modes are the one with highest angular number $\ell$; whereas, for $\bar{m}>\bar{m}_c$, the longest-lived modes are the one with smallest angular number. Also, when the overtone number $n$ increases the parameter $\bar{m}_c$ increases. Also, it is possible to observe that the behaviour of the critical parameter $\bar{m}_c$ with respect to the parameter $\bar{k}$  agrees with Fig. \ref{criticalmass2}.

\newpage

The above description of $\bar{V}_{eff}(\bar{r})$, $\bar{m}_c$, $Re(\bar{\omega})$, $Im(\bar{\omega})$ show that the scale of the location of the horizon is not important, due to the behaviour for $\bar{k}<1$ is similar to the behaviour  for $\bar{k}>1$, with $\bar{k}=k/r_h$. 

%\clearpage

\acknowledgments

We thank the referee for his/her constructive comments as
well as for useful comments and suggestions. We thank Kyriakos Destounis for carefully reading the manuscript and for his comments and suggestions. 
This work is partially supported by ANID Chile through FONDECYT Grant Nº 1220871  (P.A.G., and Y. V.). P. A. G. acknowledges the hospitality of the Universidad de La Serena where part of this work was undertaken.

\end{document}